\pdfoutput=1

\documentclass[Afour,sageh,times]{sagej}

\newif\ifarxiv
\arxivtrue

\usepackage{moreverb,url}

\usepackage[breaklinks=true, bookmarks=true, bookmarksdepth=3, colorlinks,bookmarksopen,bookmarksnumbered,citecolor=red,urlcolor=red]{hyperref}

\usepackage[T1]{fontenc}
\usepackage{graphicx}

\usepackage[backend=biber]{biblatex}
\usepackage{listings}
\usepackage{xcolor} 
\definecolor{sparqlKeyword}{RGB}{0,0,170}
\definecolor{sparqlIri}{RGB}{0,128,0}

\lstdefinelanguage{SPARQL}{
  morekeywords={SELECT,CONSTRUCT,ASK,DESCRIBE,WHERE,PREFIX,BASE,DISTINCT,REDUCED,FROM,NAMED,UNION,OPTIONAL,FILTER,GRAPH,GROUP,ORDER,BY,HAVING,LIMIT,OFFSET,DESC,ASC,AS,BIND,VALUES,SERVICE,MINUS,COALESCE,UNDEF,a},
  sensitive=true,
  morestring=[b]",
  moredelim=[s][\color{sparqlIri}]{<}{>},
}

\lstdefinestyle{sparql}{
  language=SPARQL,
  keywordstyle=\color{sparqlKeyword}\bfseries,
  stringstyle=\color{sparqlIri},
  frame=single,
  framesep=5pt,
  xleftmargin=8pt,
  xrightmargin=8pt,
  aboveskip=1em,
  belowskip=1em,
}

\graphicspath{{build/figures/}}
\usepackage{sepfootnotes}
\usepackage{amsmath}
\usepackage{stmaryrd}
\usepackage{amsfonts}
\usepackage{amssymb}
\usepackage{algorithm}
\usepackage{algorithmic}
\usepackage{subfig}
\usepackage{placeins}
\usepackage{nameref}
\usepackage{enumitem}
\usepackage{multirow}
\usepackage{amsthm}

\newtheorem{definition}{Definition}

\newif\ifanonymous

\newcommand{\mIRI}{\ensuremath{\mathit{IRI}}}
\newcommand{\miri}{\ensuremath{\mathit{iri}}}

\newcommand\BibTeX{{\rmfamily B\kern-.05em \textsc{i\kern-.025em b}\kern-.08em
T\kern-.1667em\lower.7ex\hbox{E}\kern-.125emX}}

\ifarxiv
  \def\volumeyear{\the\year}
  \def\journalname{Preprint}
\else
  \def\volumeyear{2026}
  \def\journalname{Semantic Web journal}
\fi

\usepackage{csquotes}
\usepackage{acro}
\DeclareAcronym{LTQP}{
  short = LTQP,
  long  = Link Traversal Query Processing,
}
\DeclareAcronym{HTTP}{
  short = HTTP,
  long  = Hypertext Transfer Protocol,
}
\DeclareAcronym{RDF}{
  short = RDF,
  long  = Resource Description Framework,
}
\DeclareAcronym{KG}{
  short = KG,
  long  = knowledge graph,
  long-plural-form = knowledge graphs,
}
\DeclareAcronym{IRI}{
  short = IRI,
  long  = Internationalized Resource Identifier,
  long-plural-form = Internationalized Resource Identifiers,
}
\DeclareAcronym{URI}{
  short = URI,
  long  = Uniform Resource Identifier,
}
\DeclareAcronym{DKG}{
  short = DKG,
  long  = Decentralized Knowledge Graph,
  long-plural-form = Decentralized Knowledge Graphs,
}
\DeclareAcronym{DESP}{
  short = DESP,
  long  = Decentralized Environment with Structural Properties,
  long-plural-form = Decentralized Environments with Structural Properties,
}
\DeclareAcronym{BGP}{
  short = BGP,
  long  = Basic Graph Pattern,
  long-plural-form = Basic Graph Patterns,
}
\DeclareAcronym{GSP}{
  short = GSP,
  long  = graph star pattern,
  long-plural-form = graph star patterns,
}
\DeclareAcronym{UGP}{
  short = UGP,
  long  = union graph pattern,
  long-plural-form = union graph patterns,
}
\DeclareAcronym{LDF}{
  short = LDF,
  long  = Linked Data Fragments,
}
\DeclareAcronym{TPF}{
  short = TPF,
  long  = Triple Pattern Fragments,
}
\DeclareAcronym{LDQL}{
  short = LDQL,
  long  = Linked Data Query Language,
}
\DeclareAcronym{SWSL}{
  short = SWSL,
  long  = Subweb Specifications Language,
}
\DeclareAcronym{LDP}{
  short = LDP,
  long  = Linked Data Platform,
}
\DeclareAcronym{SHACL}{
  short = SHACL,
  long  = Shapes Constraint Language,
}
\DeclareAcronym{ShEx}{
  short = ShEx,
  long  = Shape Expressions,
}
\DeclareAcronym{LDBC}{
  short = LDBC,
  long  = Linked Data Benchmark Council,
}
\DeclareAcronym{VoID}{
  short = VoID,
  long  = Vocabulary of Interlinked Datasets,
}
\DeclareAcronym{FAIR}{
  short = FAIR,
  long  = {Findable, Accessible, Interoperable, and Reusable},
}
\DeclareAcronym{PCC}{
  short = PCC,
  long  = Pearson correlation coefficient,
}

\usepackage[dvipsnames,svgnames]{xcolor} 
\usepackage[normalem]{ulem} 

\makeatletter
\font\uwavefont=lasyb10 scaled 700
\def\spelling{\bgroup\markoverwith{\lower3.5\p@\hbox{\uwavefont\textcolor{Red}{\char58}}}\ULon}
\def\grammar{\bgroup\markoverwith{\lower3.5\p@\hbox{\uwavefont\textcolor{LimeGreen}{\char58}}}\ULon}
\def\phrasing{\bgroup\markoverwith{\lower3.5\p@\hbox{\uwavefont\textcolor{RoyalBlue}{\char58}}}\ULon}

\newcommand\remove{\bgroup\markoverwith{\textcolor{red}{\rule[0.5ex]{2pt}{0.4pt}}}\ULon}
\newcommand\insertion{\bgroup\markoverwith{\textcolor{Green}{\rule[-0.5ex]{2pt}{0.6pt}}}\ULon}
\makeatother

\newif\ifreview

\colorlet{revone}{NavyBlue}      
\colorlet{revtwo}{BurntOrange}   
\colorlet{revboth}{Purple}       

\newcommand{\rone}[1]{\ifreview\textcolor{revone}{#1}\else#1\fi}
\newcommand{\rtwo}[1]{\ifreview\textcolor{revtwo}{#1}\else#1\fi}
\newcommand{\rboth}[1]{\ifreview\textcolor{revboth}{#1}\else#1\fi}

\newenvironment{Rone}{\ifreview\color{revone}\fi}{}
\newenvironment{Rtwo}{\ifreview\color{revtwo}\fi}{}
\newenvironment{Rboth}{\ifreview\color{revboth}\fi}{}

\ifreview
  \usepackage[mathlines]{lineno}
  \makeatletter
  \newcommand*\@LN@outerlinenumber{%
    \if@firstcolumn \makeLineNumberLeft \else \makeLineNumberRight \fi}
  \makeatother
\fi

\begin{document}
\ifreview
  \linenumbers
  \makeatletter\let\makeLineNumber\@LN@outerlinenumber\makeatother
\fi
\runninghead{Tam et al.}

\title{Traveling with a Map: Reducing the Search Space of Link Traversal Queries Using RDF Shapes}
\ifanonymous
      \author{Anonymous}
    \else
      \author{Bryan-Elliott Tam\affilnum{1}, Joanna Van Herwegen\affilnum{1}, Pieter Colpaert\affilnum{1}, Ruben Verborgh\affilnum{1} and Ruben Taelman\affilnum{1}}

      \affiliation{\affilnum{1}Department of Electronics and Information Systems, Ghent University – imec}

      \corrauth{Bryan-Elliott Tam}

      \email{bryanelliott.tam@ugent.be}
   \fi

\begin{abstract}
The centralization of web information raises legal and ethical concerns, particularly in social, healthcare, and education applications.  
Decentralized architectures offer a promising alternative by keeping data closer to its source, yet efficient query processing remains a significant challenge.  
\ac{LTQP} enables querying across decentralized networks, however, it often suffers from long execution times and high data transfer costs due to the large number of \ac{HTTP} requests involved.
In many scenarios, queries are highly selective with respect to the data model objects distributed across the network.
For example, in a social media application where users store heterogeneous data, a query may focus solely on the posts and comments created by users, without requiring any of their additional user information.
We refer to such queries as \emph{data-model selective}.
We propose a shape-based pruning approach that relies on \emph{shape indexes} and a \emph{query-shape subsumption} algorithm to reduce the search space and, consequently, the number of \ac{HTTP} requests for such queries.
We formalize this approach as a link pruning mechanism for \ac{LTQP} and evaluate its effectiveness on social media queries using the SolidBench benchmark across multiple evaluation metrics.
Our results show that shape-based pruning substantially improves query execution time, first-result arrival time, diefficiency, and network usage for data-model selective queries, while having a negligible impact on non-selective data-model queries.  
These gains come at the cost of only a minor increase in the number of triples per shape-index instance.  
Moreover, our approach is resilient, retaining performance benefits even in networks where some data providers do not supply shape-index information.  
This work demonstrates that shape-based metadata can significantly optimize \ac{LTQP} in decentralized knowledge graphs for an important class of queries.
By exposing such metadata, data providers not only enhance data quality and interoperability but also improve the efficiency of traversal-based query processing.  
\end{abstract}


\maketitle

\acresetall
\renewcommand{\thefootnote}{\arabic{footnote}}
\setcounter{footnote}{1}

\section{Introduction}

\sepfootnotecontent{sf:formalDataModelSelective}{
	A formal definition is provided in the \hyperref[sec:preliminaries]{preliminaries}
}

Multiple studies have highlighted that social web applications relying on highly centralized data management systems give rise to problems of ownership, threats to democracy, reinforcement of inequality, and antagonism between users and owners~\cite{Terranova2000FreeLP, Curran2016ch1, Sevignani2013, 9663788}.
\rone{Decentralizing data over the web is a response to these problems and an integral component of initiatives focused on data sovereignty, even if several authors consider it an insufficient solution on its own~\cite{9663788, Curran2016ch1}.
	Linked Data and \acp{KG}~\cite{heath2011} are technical contributions toward such a decentralized web of data.}
However, SPARQL, the standard query language for \ac{RDF} knowledge graphs, is predominantly performed in centralized environments, partly due to the more mature understanding of query optimization in such settings.
\ac{LTQP}~\cite{Hartig2012} is a query paradigm designed for querying non-indexed, \acp{DKG} on the web, by leveraging the descriptive power of \rtwo{\ac{IRI}}\footnote{\rtwo{An \ac{IRI}~\cite{ietf3987IRI} is a unique sequence of characters that identifies a resource on the web; it generalizes the \ac{URI}~\cite{ietf2396URI} by additionally permitting characters from the Unicode set.}} dereferencing.
\ac{LTQP} involves recursively dereferencing \acp{IRI}, dynamically discovering and storing triples from the documents associated with those \acp{IRI} in an internal triple store, thereby expanding the engine’s underlying knowledge base during query execution.
The main difficulty of \ac{LTQP} is the large domain of exploration, which leads to a high number of \ac{HTTP} requests as demonstrated by \citeauthor{hartig2016walking}~\cite{hartig2016walking}.
From another perspective, it has been shown that in \acp{DESP}, it is possible to attain query completeness for various types of practical queries with acceptable execution times for the context of social media applications~\cite{Taelman2023}.
\rone{Such environments are decentralized networks in which data providers expose implicit or explicit data structures, referred to as \emph{structural properties}, that query engines can leverage during traversal.}
Structural properties ensure data discoverability, which in turn helps guarantee result completeness.

In practice, \acp{DESP} emerge in various contexts, such as social networks~\cite{Taelman2023} and the publication of sensor data~\cite{tam_iswc_traversalsensortree_2024}, among others.
The work on the topic of \acp{DESP} for \ac{LTQP}~\cite{Taelman2023, Bogaerts2021LinkTW} suggests that various optimizations are feasible for \ac{LTQP} in such environments, \rone{in contrast to \citeauthor{hartig2016walking}~\cite{hartig2016walking}, who concluded that the large search space of \ac{LTQP} offers limited opportunities for optimization.}

\begin{figure*}
	\centering
	\def\svgwidth{1.0\linewidth}
	\input{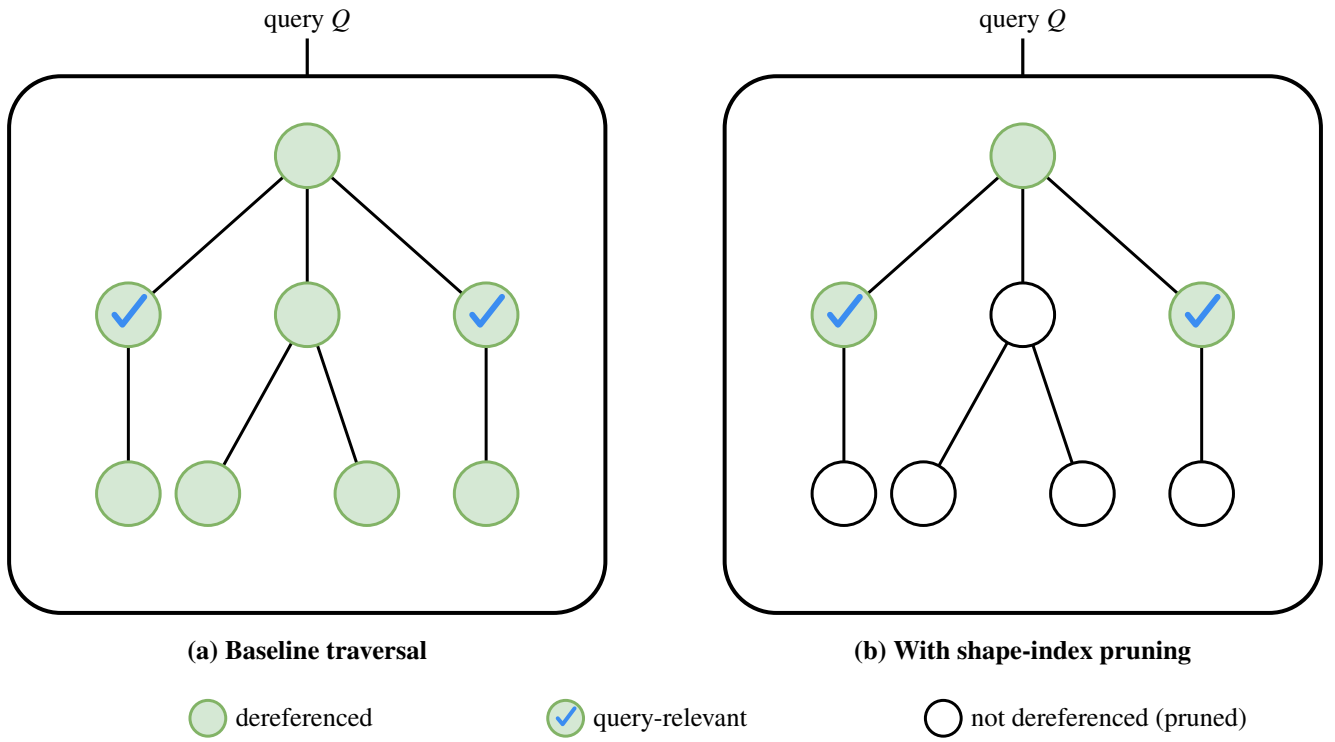}
	\caption{
		\rtwo{The same query $Q$ is issued to a network under two strategies.
		(a)~A baseline link-traversal engine dereferences many sources (green) even though only a few are query-relevant (blue check).
		(b)~Our shape-index pruning dereferences only the entry point and the query-relevant sources, leaving the sources that are provably irrelevant not dereferenced (white).
		The goal of our work is to minimize the number of query-irrelevant sources that are dereferenced by a link traversal query engine.}
	}
	\label{fig:situation}
\end{figure*}

In general, data on the World Wide Web lack a structure that query engines can exploit for optimization.
Any document can be published anywhere, with no standard index or trust mechanism to guide discovery.
Thus, for the vast majority of queries, traversal query engines dereference a large proportion of resources that are not query-relevant, as depicted in Figure~\ref{fig:situation}.
However, within specific \emph{subwebs}, that is, subsections of the web controlled by a single data provider, the data often exhibits implicit or explicit structure that query engines can exploit~\cite{Bogaerts2021LinkTW}.
In this work, we extend a dataset summarization approach for decentralized environments known as the \emph{shape index}~\cite{tam2024opportunitiesshapebasedoptimizationlink}.
We apply this approach to enable link pruning within \ac{LTQP}, removing links that are not relevant to the query, based on an analysis of \ac{RDF} data shapes and the user's query.
\rone{Whereas existing traversal methods focus on discovering which links to follow, our method prunes links that can be proven irrelevant, removing even a discovered link when the data shapes show it cannot contribute to the query results.}
The analysis is performed by conducting a query-shape subsumption check to determine whether a resource conforming to a given shape is relevant to a query.
Our approach assumes a \ac{DKG} composed of subwebs, each hosted by data providers and containing shape indexes.
A subweb that exposes a shape index enables the query engine to narrow its search domain by identifying resources that are certainly non-query-relevant.
\textbf{This is particularly useful when only a subset of a subweb is relevant to a given query, for example, in social media applications where it is rare to query all information about a user.}
An example of such a query is one that requests messages within a user's subweb over a specified date range, while also incorporating contextual information from outside the subweb.
We describe this category of network-tied queries as \emph{data-model selective}.
In this context, \emph{data-model objects} are templates of concepts represented in \ac{RDF},
such as messages, posts, or proteins.
Accordingly, we consider \ac{RDF} data shapes to be formal descriptions of these data-model objects.
For the purposes of this paper, we define a \emph{data-model selective query}
as a query whose evaluation requires fewer than 20\% of the data-model objects within the network.
In this work, we do not aim to optimize queries that retrieve a large proportion of a subweb's data model objects.
For instance, queries that request all user information published after a given date are outside the scope of our optimization; however, we attempt to ensure that the performance of such queries is not degraded.

\begin{Rboth}
	Our contributions are threefold:
	\begin{description}
		\item[Method.] A novel shape-based pruning method for \ac{LTQP} that introduces the following new notions:
		      \begin{enumerate}[label=(\roman*)]
			      \item \emph{result-based completeness}, formalized as a search-space minimization problem;
			      \item \emph{composite reachability criteria} that combine link discovery and pruning;
			      \item a network abstraction of \emph{subwebs} and \acp{DKG}; and
			      \item \emph{shape-based link pruning}, which uses \ac{RDF} data shapes to discard sources that cannot contribute to the query results while preserving result completeness.
		      \end{enumerate}
		\item[Software components.] Three artifacts\footnote{\rtwo{The shape index specification and the open-source implementations of all software components are available online: \ifanonymous\url{https://anonymous.4open.science/w/shape-index-specification-707E} and \url{https://anonymous.4open.science/r/documentation-1A65}\else\url{https://constraintautomaton.github.io/shape-index-specification/} and \url{https://github.com/shapeIndexComunicaExperiment/documentation}\fi.}} realize the method:
		      \begin{enumerate}[label=(\roman*)]
			      \item a web specification for the \emph{shape index} data summary;
			      \item a \emph{query-shape subsumption algorithm} for assessing data source relevance;
			      \item an open-source implementation in the Comunica SPARQL query engine~\cite{taelman_iswc_resources_comunica_2018}.
		      \end{enumerate}
		\item[Evaluation.] An extensive evaluation on the SolidBench benchmark, across multiple metrics (query execution time, first-result arrival time, diefficiency, and number of \ac{HTTP} requests), comprising:
		      \begin{enumerate}[label=(\roman*)]
			      \item a comparison against state-of-the-art \ac{LTQP} traversal (the type index and \ac{LDP} approaches), showing that shape-based pruning reduces the number of \ac{HTTP} requests and improves query execution time for data-model-selective queries, and even enables a query that no prior approach could complete, while leaving non-selective queries essentially unaffected;
			      \item a cost analysis of the query-shape subsumption algorithm, showing that its overhead is negligible relative to query execution;
			      \item a resilience study under partial and reduced shape-index coverage, showing that the performance gains persist even when not all data providers expose a shape index.
		      \end{enumerate}
	\end{description}
\end{Rboth}

In this work, we ask the following research question:
\textbf{Can \ac{LTQP} use shape-based pruning in \ac{DKG} networks to reduce query execution time while preserving result completeness?}
\begin{Rone}
	To address this question, we formulate hypotheses along the four axes that structure our evaluation:
	\begin{description}
		\item[Effectiveness.] Whether pruning improves performance for the targeted queries without harming the others:
		      \begin{itemize}
			      \item[\textbf{H1}] The shape indexes approach reduce the number of non-contributing data sources retrieved and the query execution time of \emph{data-model selective} queries, while preserving result completeness.
			      \item[\textbf{H2}] The shape indexes approach does not affect the query execution time of \emph{non-data-model selective} queries.
		      \end{itemize}
		\item[Cost.] Whether the pruning mechanism is itself inexpensive:
		      \begin{itemize}
			      \item[\textbf{H3}] The execution time of a query-shape subsumption algorithm is negligible compared to the query execution.
		      \end{itemize}
		\item[Robustness.] How the amount of shape-index information across the network affects the performance:
		      \begin{itemize}
			      \item[\textbf{H4}] Stricter shape constraints lead to a greater reduction in \ac{HTTP} requests.
			      \item[\textbf{H5}] Querying a network with more \emph{complete} shape indexes results in faster query execution.
			      \item[\textbf{H6}] Performance gains can be achieved even in networks with partial shape index coverage, demonstrating that the method provides significant improvements without requiring all network participants to adopt shape indexes.
		      \end{itemize}
		\item[Explainability.] What relationship explains the observed gains:
		      \begin{itemize}
			      \item[\textbf{H7}] There is a linear relationship between the reduction in the number of \ac{HTTP} requests and the decrease in query execution time.
		      \end{itemize}
	\end{description}
\end{Rone}

The remainder of this paper is organized as follows:
we first review the \hyperref[sec:related_work]{related work} and introduce the \hyperref[sec:preliminaries]{preliminaries}.
Next, we present our \hyperref[sec:approach]{approach}, followed by a description of the \hyperref[sec:experiment]{experimental setup} and a \hyperref[sec:result]{discussion of the results}.
Finally, we \hyperref[sec:conclusion]{conclude the paper}.

\section{Related Work}\label{sec:related_work}

In this section, we review related work on \ac{LTQP}, \ac{RDF} data shapes, and source selection in decentralized querying.

\subsection{Link Traversal Query Processing}
\ac{LTQP} is a SPARQL querying paradigm that answers queries by exploring the Web using the \textit{follow-your-nose} principle\rone{, whereby the engine dereferences the \acp{IRI} found in retrieved documents to discover further documents to query}~\cite{hartig2016walking}.
It belongs to the family of decentralized SPARQL querying paradigms.
\ac{LTQP} fundamentally differs from federated querying because the ``federation'' is formed during querying and it expands dynamically as the query is processed.
\rone{A federation is a set of queryable \ac{RDF} sources, typically SPARQL endpoints, that is fixed and known prior to query execution.}
Thus, many optimization techniques used in federated querying either do not work in the context of \ac{LTQP} or remain unexplored.
\ac{LTQP} also differs from querying \ac{LDF} interfaces~\cite{azzam2020smart,azzam2021wisekg,DBLP:journals/corr/HartigA16}, as \ac{LDF} operates over fragments of a known, local, finite \ac{KG}, whereas \ac{LTQP} dynamically discovers new, remote \acp{KG} during query processing.
In summary, \ac{LDF} methods such as \ac{TPF}~\cite{verborgh2016triple} focus on the efficient partitioning of a known dataset, while \ac{LTQP} focuses on the efficient discovery of an unknown dataset.
\rone{More generally, an \ac{LDF} interface publishes a dataset as a set of queryable fragments over the Web, balancing query-processing effort between the server and the client~\cite{verborgh2014web}.}

The main challenge of \ac{LTQP} is the Web's open-ended nature leading to large search spaces.
Completeness in \ac{LTQP} is defined by the traversal of a well-defined set of links~\cite{Hartig2012}.
\rone{A traversal is complete when it dereferences every link in this set, so that the query is answered completely over the subgraph those links induce; completeness in \ac{LTQP} is therefore relative to the set of reachable links rather than to the entire Web, whose search space is too large to traverse exhaustively.}
The first method used to define this set was the \emph{reachability criteria}~\cite{Hartig2012}, boolean functions that determine whether a given link should be dereferenced.
In practice, the query engine iterates over all triples in its internal data source and applies the reachability criteria to each \ac{IRI} appearing in those triples.
These criteria are defined internally by the engine, giving designers considerable freedom in how they are implemented and which links are dereferenced. 
However, there is currently no standard mechanism for users to specify them, nor for query engine developers to adopt a common approach.
Building on this, the theoretical query language \ac{LDQL}~\cite{hartigLDQL} was introduced, which separates the traversal definition from the query definition.
It provides a standard mechanism for query engines to define their traversal logic, while opening the possibility for users to provide a traversal policy alongside their queries.
Further advancements include the \ac{SWSL}~\cite{Bogaerts2021LinkTW}, which allows data providers to define how their \ac{DKG} should be traversed.
Inspired by \ac{SWSL}, traversal-based querying has utilized the \ac{LDP} and the Type Index specification~\cite{Taelman2023}.
\ac{LDP} traversal follows all links within a data space, while Type Index traversal uses mappings from \ac{RDF} types to relevant resources~\cite{solidTypeIndexes}, allowing queries to prioritize implicitly relevant data sources.
Link prioritization has also been studied, however, it has been shown to not improve query execution time~\cite{hartig2016walking}, even in \acp{DESP}~\cite{eschauzier_iswc_research_linkprioritization_2025}.
These contributions are centered on guiding the engine in selecting links to follow in a discovery process.
However, they do not explicitly address the restriction or pruning of links after the discovery process based on information acquired during traversal.
\rone{We distinguish discovery from pruning as assertion differs from negation: a discovery criterion expresses that a link should be explored, whereas a pruning criterion expresses that it should not, and the latter takes precedence, so a pruned link is not followed even when a discovery criterion would select it.
Discovery builds the search domain of an \ac{LTQP} query, whereas pruning carves holes into that domain.
Although both can be described as retaining a subset of links, only inclusion is sound under the open-world assumption of the web, since a missing link is no evidence that relevant data is absent; excluding links soundly requires closed-world knowledge about a region of the web.}
Such pruning could significantly reduce the query search domain when information about the \ac{DKG} data model is available.
For instance, the structural properties of a subweb could inform the query engine that certain web sections follow a specific data model, allowing a set of \acp{IRI} to be pruned from those selected during discovery.
In contrast, when \ac{LTQP} models \acp{DKG} as Linked Open Data, the web is not divided into subsections with structural properties; thus, data model information cannot be inferred.
To the best of our knowledge, no prior work has explored the use of a pruning mechanism to optimize \ac{LTQP}, and it is this research gap that the present study aims to address.

\subsection{RDF Data Shapes}
\ac{RDF} data shapes (in this paper, we also refer to them as shapes) are used for validating, describing, and communicating data structures, as well as generating data and driving user interfaces~\cite{Gayo2018a,Gayo2018}.
The two most well known \ac{RDF} data shape formalisms are \ac{SHACL}~\cite{Gayo2018b} and \ac{ShEx}~\cite{Gayo2018}.
For common use cases, they are equally expressive and interchangeable~\cite{Gayo2018c}.
\ac{RDF} data shapes have already been used in the literature for querying centralized \acp{KG}~\cite{kashif2021, delva2023}.
Shape Trees~\cite{shapetreesShapeTrees} are an index structure for validating and organizing \acp{DKG}; \rone{however, to the best of our knowledge, they have not been used for query optimization, and their \emph{virtual hierarchy}~\cite{shapetreesShapeTrees} makes it challenging for a query engine to efficiently capture the relationship between a resource \ac{IRI} and its corresponding shape.
Beyond this conceptual mismatch, Shape Trees were not yet practical to build upon at the time of this study: the specification is not widely adopted, and its supporting documentation and tooling were not sufficiently mature or stable.
We therefore conduct our investigation using the Shape Index specification~\cite{tam2024opportunitiesshapebasedoptimizationlink} to map shapes to \acp{KG}.
Importantly, our findings are not tied to this particular summary: our approach relies in essence on a mapping between shapes and sets of \acp{IRI}, so any structure providing such a mapping---including a more mature Shape Trees---could serve as the underlying summary instead.}
Additionally, automatic generation of \ac{RDF} data shapes from \acp{KG}~\cite{fernandez2023extracting} and shape-based data integration~\cite{LabraGayo2023} have been studied and can support shape-based summary approaches for \acp{DKG}.

\subsection{Source Selection}
Source selection is a crucial challenge in decentralized querying~\cite{hose2012towards, Harth2010}.
\rone{It is the process of determining, before or during query execution, which of the available data sources can contribute to the result of a query, so that sources that cannot are not accessed.}
Link pruning in \ac{LTQP} is closely related to source selection, as it can be viewed as a dynamic form of source selection.
Methods such as basic statistics on triple counts, \ac{VoID} descriptions, and histogram techniques have been explored in the context of federated querying~\cite{hose2012towards, Harth2010, Montoya2017}.
\rone{A \ac{VoID} description is an \ac{RDF} document that summarizes a dataset, reporting, among other things, its total number of triples and the set of predicates it uses. Figure~\ref{fig:void} shows such a description for the \texttt{sw:posts} dataset of our running example.}

\begin{figure}
   \centering
\begin{lstlisting}[basicstyle=\footnotesize\ttfamily]
@prefix void: <http://rdfs.org/ns/void#> .
@prefix ex:   <http://example.com/> .
@prefix sw:   <http://subweb/> .

sw:posts a void:Dataset ;
    void:entities 160 ;
    void:triples 480 ;
    void:propertyPartition [
        void:property ex:creator ;
        void:triples 160 ] ;
    void:propertyPartition [
        void:property ex:content ;
        void:triples 160 ] ;
    void:propertyPartition [
        void:property ex:reply ;
        void:triples 160 ] .
\end{lstlisting}
   \caption{\rone{A short \ac{VoID} description of the \texttt{sw:posts} dataset from the running example of Figure~\ref{fig:shapeIndex}, describing $160$ posts that together amount to $480$ triples, partitioned over the \texttt{ex:creator}, \texttt{ex:content}, and \texttt{ex:reply} predicates.}}
   \label{fig:void}
\end{figure}
However, most of those source selection methods face the limitation of assuming a small number of data sources~\cite{Harth2010}, leaving their suitability for \ac{LTQP} uncertain.
Bloom filters~\cite{dia2018fast} are also a mechanism that has shown success for federated \acp{DKG}, yet in the context of \ac{LTQP}, it has been shown that bloom filters have little effect on performance~\cite{Hanski2024}.
Schema-based indexing using ontologies~\cite{Stuckenschmidt2004} has also been explored for source selection of SPARQL queries.
It has been shown that this approach is sensitive to the high reuse of vocabulary terms in \ac{RDF}~\cite{Harth2010}, which is exacerbated in the context of \ac{LTQP}.
The use of implicit \ac{RDF} schemas for query optimization has been explored through the concept of characteristic sets~\cite{Neumann2011CharacteristicSA, Meimaris2017ExtendedCS, Montoya2017}.
However, their applicability to \ac{LTQP} has not been investigated, and they assume that the entire dataset resides in memory, which is not the case for \ac{LTQP}.

\section{Preliminaries}\label{sec:preliminaries}

\subsection{RDF Knowledge Graphs and SPARQL Queries}

Our work focuses on the union of conjunctive queries over \ac{RDF} \acp{KG} using the SPARQL query language~\cite{w3SPARQLQuery}.
The fundamental building blocks of \acp{KG} and SPARQL queries are triples and triple patterns, respectively, as defined in Definition~\ref{def:triple} and Definition~\ref{def:triplePattern}.

\begin{definition}[Triple]\label{def:triple}
	\ac{RDF} triples $t = (s,p,o)$ are tuples formed with three terms. A subject where $s \in \mathcal{I} \cup \mathcal{B}$, a predicate $p \in \mathcal{I}$ and an object $o \in \mathcal{I} \cup \mathcal{B} \cup \mathcal{L}$.
	Where $\mathcal{I}$, $\mathcal{B}$, and $\mathcal{L}$ are respectively the set of every possible \ac{IRI}, blank node, and literal.
	For simplicity, we denote the union of these sets by concatenating their symbols, so that $ \mathcal{I} \cup \mathcal{B}$ is written as $ \mathcal{I} \mathcal{B}$.
\end{definition}

\begin{definition}[Triple pattern]\label{def:triplePattern}
	Triple patterns $tp = (s_{tp}, p_{tp}, o_{tp})$ are similar to triples, \rone{except that any of the three positions may additionally be a variable from $\mathcal{V}$, the set of every possible variable,} where $s_{tp} \in \mathcal{I} \mathcal{B} \mathcal{V}$,
	$p_{tp} \in \mathcal{I} \mathcal{V}$ and an object term  $o_{tp} \in \mathcal{I} \mathcal{B} \mathcal{V} \mathcal{L}$.
\end{definition}

We also define two access functions to respectively get the subject and object term of a triple or a triple pattern while ignoring literals,
$ S:  (\mathcal{I} \mathcal{B} \mathcal{V}, \mathcal{I} \mathcal{V}, \mathcal{I} \mathcal{B} \mathcal{V} \mathcal{L}) \rightarrow \mathcal{I} \mathcal{B} \mathcal{V}$ and $O: (\mathcal{I} \mathcal{B} \mathcal{V}, \mathcal{I} \mathcal{V}, \mathcal{I} \mathcal{B} \mathcal{V} \mathcal{L}) \rightarrow \mathcal{I} \mathcal{B} \mathcal{V}$\rone{, such that for a triple or triple pattern $(s, p, o)$, $S((s,p,o)) = s$ and $O((s,p,o)) = o$ when $o \in \mathcal{I}\mathcal{B}\mathcal{V}$, while $O$ returns nothing when $o$ is a literal}.
We denote $[\![ Q ]\!]^{G}$ as the evaluation of a query $Q$ over a \ac{KG} $G$~\cite{Angles2008}.

\subsection{Reachability Criteria}

\ac{LTQP} defines completeness on the traversal of links instead of the query results~\cite{Hartig2012}.
To formalize the completeness of queries, \emph{Reachability criteria}~\cite{Hartig2012} have been formalized.
Reachability criteria are boolean functions ($c_i$) restricting the dereferencing of links from the internal data source of the query engine.
\rtwo{They take three parameters: an \ac{RDF} triple $t$ from an internal triple store, a dereferenceable \ac{IRI} $iri$ occurring in $t$, and a union of conjunctive queries $Q$.}
\rone{Their domain and range are $c_i: \big((\mathcal{I}\mathcal{B}, \mathcal{I}, \mathcal{I}\mathcal{B}\mathcal{L}),\ \mathcal{I},\ \mathcal{Q}\big) \rightarrow \left\{ \mathrm{true}, \mathrm{false} \right\}$, where $\mathcal{Q}$ is the set of unions of conjunctive queries.}
If $c_i$ returns $true$, the query engine must dereference $iri$.
\rone{Dereferencing an \ac{IRI} means issuing an \ac{HTTP} request to it to retrieve a resource.}
More formally
\begin{equation}\label{eq:reachabilityCriteria}
	c_i(t, iri, Q) \rightarrow \left\{ \mathrm{true}, \mathrm{false} \right\}
\end{equation}

\subsection{Decentralized Knowledge Graphs and Subweb}\label{sec:dkg}

We define a \ac{DKG} as a \rone{single} \ac{KG} $G$ materialized in a network of resources $R$.
A resource $r_i \in R$ \rone{is a document that materializes exactly one} \ac{KG} $g_i \subseteq G$, which is a set of triples~\cite{w3ConceptsAbstract}.
We denote this mapping $r_i \mapsto_{\mathcal{G}} g_i$.
\rone{A resource is exposed by an \ac{IRI} $iri_i$, an identifier pointing to it, denoted by $iri_i \mapsto_{\mathcal{R}} r_i$; the resource is thus distinct from the \ac{IRI} that identifies it.}
The network forms a graph where the resources $r_i$ are the nodes and the $iri_j$ are directed edges starting from $r_i$ to $r_j$.
The $iri_j$ are \ac{RDF} terms in the triples in $g_i$.
$G$ is formed by the union of all the $g_i$ mapped to a resource in the network.
A subweb is a (sub)\ac{DKG} defined by the \ac{KG} derived from a set of \acp{IRI} controlled by a data provider.

\subsection{Data-model Selectiveness}
Data-model selectiveness is an ordering of queries over networks.
A \emph{data-model object} is a well-defined template of sub-\acp{KG} \rone{that corresponds to a single data type, such as a user or a post}.
By \emph{well-defined}, we mean that there exist explicit rules which deterministically specify the constraints on the triples within the data-model object, as well as its boundary within a larger \ac{KG}.
In this work, we assume that data-model objects are described using \ac{RDF} data shapes, however, they could also be described using vocabularies or other formalisms.
We define $DM$ as the set of the data-model objects present in the networks.
We define a function to determine all the relevant data-model objects for a given query:
\begin{equation}
	P(Q) = \{ D \mid (D \in DM) \text{ \textbf{is query relevant for} } Q \}.
\end{equation}
\rone{A data-model object $D$ is query relevant for $Q$ when at least one conjunct of $Q$ can be answered by a \ac{KG} derived from $D$.}
A query $Q_i$ is more data-model selective than another query $Q_j$ if and only if
\begin{equation}
	\left| P(Q_i) \right| < \left| P(Q_j) \right|.
\end{equation}

For practical use cases, we may consider only the data-model objects present in a single subweb $s$ rather than the whole \ac{DKG}, \rone{replacing $DM$ by $DM^s$, the set of data-model objects in subweb $s$, and $P$ by the corresponding function $P^s(Q)$}.
Furthermore, for practical designation, we say that a query is \emph{data-model selective} (in absolute terms) if
\begin{equation}
	\frac{\left| P^s(Q) \right|}{\left| DM^s \right|} \leq 0.2,
\end{equation}
that is, if at most 20\% of the data-model objects are potentially relevant to the query.\footnote{\rone{The $0.2$ threshold is arbitrary. There is no objective criterion for choosing such a value, so we picked it solely for this practical designation.}}

\sepfootnotecontent{sf:shapeIndexURL}{
	\ifanonymous
		\url{https://anonymous.4open.science/w/shape-index-specification-707E}
	\else
		\url{https://constraintautomaton.github.io/shape-index-specification/}
	\fi
}

\sepfootnotecontent{sf:recursiveShape}{
	We only consider shapes that can be transformed into a single \texttt{SELECT} SPARQL query~\cite{Corman2019}.
}

\section{Approach}\label{sec:approach}

This section defines result-based completeness in \ac{LTQP}, introduces shape indexes, and shows how they enable pruning via a query-shape subsumption algorithm.

\begin{figure*}
	\centering
	\includegraphics[width=1.0\linewidth]{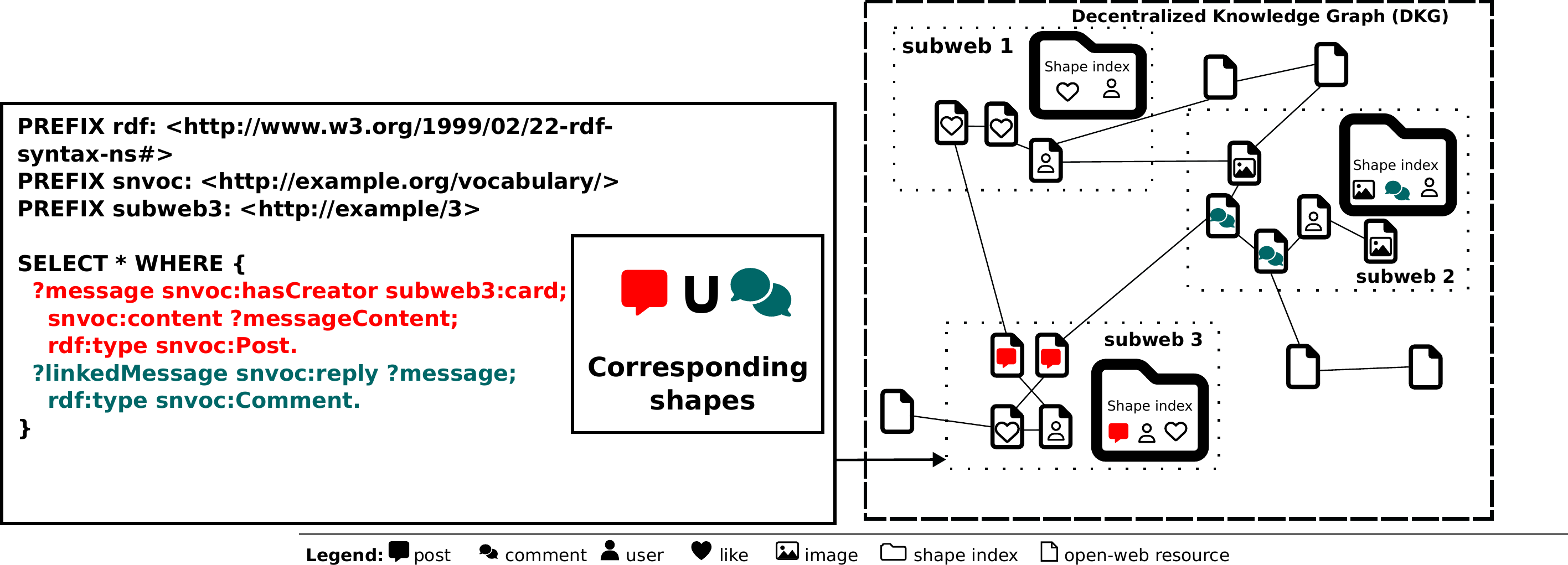}
	\caption{
		When resources of a \ac{DKG} are indexed with a shape index, a query engine can dereference a subset of the network.
		\rboth{Each node is an \ac{RDF} resource, a document identified by an \ac{IRI}, and each edge is an \ac{IRI} in one resource that links to another.
			The icon embedded in a node indicates the shape, and thus the data type such as a post, comment, or user, that the resource conforms to; the legend at the bottom of the figure defines these icons.
			Each subweb publishes a shape index, drawn as a folder, that maps the shapes its resources conform to, so Subweb~1 contains user and like resources, Subweb~2 user, comment, and image resources, and Subweb~3 post, user, and like resources.
			Documents outside the subwebs are resources from the open web and are thus not bound by any shape.}
		The query engine starts its query at Subweb~3\rboth{, and the colored documents are those relevant to the query shown on the left: red for posts and green for comments}.
	}
	\label{fig:dkg}
\end{figure*}

To illustrate this approach, we present the example in Figure~\ref{fig:dkg}, which depicts a network of three social media user subwebs, each with its own shape index, as well as resources located outside these subwebs.
The feature query aims to retrieve posts from Subweb~3, along with all associated replies.
Our pruning strategy allows the query engine to explore only the relevant parts of the network, guided by the shapes associated with the resources and the structure of the query.
The process begins with the engine dereferencing the shape index of Subweb~3 and performing a query-shape subsumption check, which determines that only resources containing posts need to be accessed in this subweb.
It then checks the shape index of Subweb~1 due to the existing link towards it, where the subsumption check reveals no resources relevant to the query.
Next, it examines the shape index of Subweb~2 and identifies resources containing comments (i.e., replies) which are relevant.
Finally, the engine dereferences all reachable resources outside the subwebs that are linked to these relevant comments.
\rone{Figure~\ref{fig:shapeIndex} illustrates an example of a shape index alongside a query pattern that our approach analyzes; both are formalized in the following subsections.}

\subsection{Result-Based Completeness in LTQP}\label{sec:slde}

Our approach of pruning in \ac{LTQP} focuses on ensuring result completeness, assuming traversal completeness is already defined using reachability criteria.
By concentrating on result completeness, we explore strategies to optimize the search space of link traversal queries through pruning of irrelevant resources.
We formalize result-based completeness in \ac{LTQP} as follows.
A query is executed over a \ac{DKG} $G$ formed by the union of all the $g$ in a network $R$.
The query engine has to build a \ac{KG} $G^{\prime}$ using a reachability criterion $C^{\prime}$ in its internal data store from the \acp{KG} $g$ by dereferencing resources $iri \mapsto_{\mathcal{R}} r \in R$.
We formulate an optimization problem to minimize the size of $G^{\prime}$, where the query engine constructs a \ac{KG} $G^{\prime\prime} \subseteq G^{\prime}$, potentially smaller, by defining a reachability criterion $C^{\prime\prime}$.
We focus on maintaining the same result completeness, so when using $C^{\prime\prime}$ the following equation must hold

\begin{equation}\label{eq:evalQueryStructuralAssumption}
	[\![ Q ]\!]^{G^{\prime\prime}} = [\![ Q ]\!]^{G^{\prime}}
\end{equation}
for any network $R$.
Since each $g \in G^{\prime\prime}$ is obtained by dereferencing resources $r \in R$, a smaller $G^{\prime\prime}$ compared to $G^{\prime}$ implies that fewer \ac{HTTP} requests were needed to answer the query.
Query execution is generally faster with a smaller \ac{KG} instance, and \ac{HTTP} requests, being slow and unpredictable~\cite{hartig2016walking}, can dominate execution time.
Therefore, reducing \ac{HTTP} requests provides a twofold benefit: fewer resources to process and faster query execution.

\rone{
	\paragraph{Example}
	In Figure~\ref{fig:dkg}, the feature query starts at the seed document of Subweb~3.
	The discovery criterion $C^{\prime}$ follows every edge reachable from that document and therefore traverses the whole network ($17$ links).
	The more selective $C^{\prime\prime}$ follows only the edges leading to posts and their replies, traversing a subset of the network ($4$ links).
	The two criteria build different graphs $G^{\prime}$ and $G^{\prime\prime}$, yet report the same answer to $Q$ (Equation~\ref{eq:evalQueryStructuralAssumption} holds), needing only $4$ dereferences instead of $17$.
}

\subsection{Shape Index}

\begin{figure*}
	\centering
	\begin{minipage}[t]{0.50\linewidth}
		\centering
\begin{lstlisting}[basicstyle=\normalsize\ttfamily]
@prefix sw: <http://subweb/> .
@prefix si: <https://shapeindex.com#>.

sw:shapeIndex si:shapeIndexLocation si:shapeIndex;
    a si:ShapeIndex;
    si:entry _:user;
    si:entry _:Post.
_:user si:shape sw:user_shape;
    si:subweb "http://subweb/user/{info}".
_:Post si:shape sw:post_shape;
    si:subweb sw:posts.
\end{lstlisting}
	\end{minipage}
	\begin{minipage}[t]{0.45\linewidth}
		\centering
\begin{lstlisting}[basicstyle=\normalsize\ttfamily]
PREFIX ex: <http://example.com/>

SELECT * WHERE {
  ?comment ex:creator ex:user;
    ex:content ?content;
    ex:reply ?message.
  ?message ex:hasTopic ?topic;
    ex:forum ?forum.
  ?forum ex:name ?name.
}
\end{lstlisting}
	\end{minipage}

	\caption{
		On the left, an example illustrates a shape index that maps a set of \acp{IRI}, represented using a \ac{URI} template~\cite{ietf6570Template}, to a user shape, and a specific \ac{IRI} to a post shape.
		On the right, an example illustrates a \ac{GSP} where the main subject is \texttt{?comment} and is linked to the \texttt{?message} and \texttt{?forum} star patterns.}
	\label{fig:shapeIndex}
\end{figure*}

Pruning in \ac{LTQP} requires knowledge of the data models of dereferenced resources.
However, obtaining complete, up-to-date, and detailed information for each resource in a large decentralized network is impractical.
To address this, we introduce the \emph{shape index} as a mapping between \ac{RDF} document sets and \ac{RDF} data shapes that describes a subweb controlled by a data provider.
Unlike triple statistics, shapes are independent of the \ac{KG}'s size or updates that remain compliant, making them a more cost-effective solution for use cases with stable data models.

We formalize a shape index as follows:

\begin{equation}\label{eq:shapeIndex}
	SI = \left\{ s_1 \mapsto \mIRI_1, s_2 \mapsto \mIRI_2, \cdots, s_n \mapsto \mIRI_n \right\}
\end{equation}
where $s_i$ is a shape and $\mIRI_i$ is a set of \ac{IRI} given $n$ entries.
The subweb described by the index is defined by $D_{SI} = \bigcup_{\mIRI \in \text{codomain}(SI)} \mIRI$.
We denote a shape index as \emph{complete} when every shape $s_i \in \text{dom}(SI)$ has a closed world assumption~\cite{Gayo2018b, Gayo2018}~\footnote{We also refer to those shapes as closed.} or \emph{incomplete} otherwise.
A mapping between a shape and a set of \acp{IRI} has implications in the distribution of the data in $D_{SI}$.
When a shape $s$ is mapped to an $\mIRI$, then the \ac{KG} targeted by the mapping, $G = \left\{ g \mid \forall iri \in D_{SI} (\miri \mapsto_{\mathcal{R}} r \land r \mapsto_{\mathcal{G}} g) \right\}$, satisfies $s$.
Given that the shape is closed, then every set of triples in the resource mapped to an $\miri \in D_{SI}$ satisfying the shape must be in a resource mapped to an $\miri \in IRI$.
We provide a complete description of the shape index in an online specification~\sepfootnote{sf:shapeIndexURL} and an example of serialization in Figure~\ref{fig:shapeIndex}.

\ac{RDF} data shapes use \emph{targets} to identify the set of nodes or entities in a \ac{KG} to validate.
In this work, we assume all entities in a \ac{KG} associated with a shape index follow the same \ac{RDF} data shape.
We call these entities graph stars, an extension of the \ac{RDF} star patterns concept~\cite{Karim2020}.
Defined in Definition~\ref{def:starPattern}, graph star(s) (patterns)  serve two purposes:
defining targets for validation and capturing relationships between triple patterns and shape entities.
Star patterns consist of triples with the same subject.
We extend this concept by linking star patterns such that the objects of triples in one star pattern act as subjects in others, forming a graph structure.
For example, a user linking to their posts with recursive replies can be captured with a root star pattern for the user and nested patterns for the posts and replies.
Thus, the targets of the shapes in the shape index correspond to the subject of each root star pattern when a \ac{KG} is divided into graph stars with no shared partial graph stars.
Figure~\ref{fig:shapeIndex} illustrates an example of a \ac{GSP}.

\begin{definition}[Star Pattern]
	Given a query $Q$, we define a star pattern $Q_{star}$ as a set of $tp \in Q$~\cite{Karim2020} with the same subject such that
	given a builder function
	\begin{equation}
		BQ_{star}(s) = \left\{ tp_i \in Q \mid S(tp_i) = s \right\}
	\end{equation}
	with $s \in \mathcal{I}\mathcal{B}\mathcal{V}$ then $Q_{star} = BQ_{star}(s)$.

\end{definition}

\begin{definition}[Graph Star Pattern (GSP) and Graph Star]\label{def:starPattern}

	We define a \ac{GSP} $Q_{starG}$ as the union between a root star pattern $Q_{star_s}$
	and the star patterns having as subject term an object term of another star pattern in $Q_{starG}$.
	We define a function
	$O_{star}: q \in Q \rightarrow  (\mathcal{I}\mathcal{B}\mathcal{V})^2$
	that returns every non-literal object terms of a star pattern.

	We then define $Q_{starG}$ given a  set of partial \acp{GSP} $Q_{starGT}$
	\begin{equation}
		Q_{starG} = \bigcup_{q \in Q_{starGT}} q
	\end{equation}
	where $Q_{starGT}$ is formed with a root $Q_{star_s}$ by

	\begin{equation}
		q_i =
		\begin{cases}
			\left\{ Q_{star_s} \right\}                                                           & \text{if } i = 1 \\
			\left\{ BQ_{star}(o) \mid o \in \bigcup_{q \in Q_{starGT_{i-1}}} O_{star}(q) \right\} & \text{if } i>1
		\end{cases}
	\end{equation}

	We also define a function
	$S_{star}: q \rightarrow  \mathcal{I}\mathcal{B}\mathcal{V}$
	returning the subjects of the $q_i$ of a $Q_{starG}$.

	We propose a similar definition for the context of \acp{KG} where we replace the query $Q$ by a \ac{KG} $G$.
	We denote this structure a graph star.

\end{definition}

\subsubsection{Practical Considerations for Shape Indexes}

The construction and maintenance of shape indexes are beyond the scope of this work.
Although not evaluated here, shape indexes seem to require less effort to generate than \ac{VoID} descriptions~\cite{Boehm2011}, as they do not include detailed statistics such as triple counts.
Nonetheless, \ac{VoID} descriptions have been successfully employed for query optimization in federated queries~\cite{Montoya2017}.

Exposed schemas can also enhance interoperability, which is important in many application domains~\cite{rowland2022interoperability, solidSolidApplication, binding2016improving, roller2015benefits}.
Thus, data publishers may have incentives to expose a shape index not only to improve query engine performance but also to satisfy other domain-specific requirements.

\ac{RDF} data shapes can be prescriptive or descriptive.
For descriptive shape indexes, automatic \ac{RDF} data shape generation methods~\cite{fernandez2023extracting} can facilitate their creation.
Entries in shape indexes correspond to sets of \acp{IRI}, which can be structured using \ac{URI} templates~\cite{ietf6570Template}, reducing the need for exhaustive redefinition.
For prescriptive shapes, contributions in shape-based data integration~\cite{LabraGayo2023} can help prevent the generation of invalid resources.

Overall, shape indexes are lightweight metadata that can be maintained with relatively low effort, especially compared to statistical summaries such as \ac{VoID}.
Publishing them not only benefits query optimization but also contributes to \ac{FAIR} data principles~\cite{Wilkinson2016} by enhancing machine-actionability and interoperability.
Future work should investigate construction and maintenance strategies for shape indexes in practical deployment scenarios.

\subsection{Link Pruning Using Shape Indexes}\label{sec:sourceSelection}

In this section, we establish the connection between shape indexes and link pruning in \ac{LTQP} as a means to reduce the search domain.
In our method, rather than traversing the entire \ac{DKG} $D$ associated with a shape index, the engine traverses a subgraph $D^{\prime} \subseteq D$, effectively ignoring resources that are knowably irrelevant to the query.
Our approach involves dynamically constructing new reachability criteria during traversal that are more selective as we discover and analyze shape indexes.
These criteria are designed so that they will always produce the same completeness of results as the one that was defined at the beginning of the traversal.

To define more selective reachabilities, we propose extending the reachability criteria by formalizing a chain of criteria in a concept called \emph{composite reachability criteria}.
In this form, a reachability criterion $cp_i$ is said to \emph{prune} links, and $cd_i$ is said to \emph{discover} links.
Equation~\ref{eq:cReachabilityCriteria} formalizes a composite reachability criterion $C$.
where $Cd$ is the set of every $cd_i(t, iri, Q)$ and $Cp$ the set of every $cp_i(t, iri, Q)$ used by the engine.

\begin{equation}\label{eq:cReachabilityCriteria}
	C(t, iri, Q) = \bigvee_{cd \in Cd} cd(t, iri, Q) \mathrel{\land} \bigwedge_{cp \in Cp} \, cp(t, iri, Q)
\end{equation}

\rone{This structure reflects the distinction between discovery and pruning: discovery criteria are combined disjunctively, so any one of them suffices to assert that a link is reachable, whereas pruning criteria are combined conjunctively, so a single pruning criterion blocks a link regardless of how many discovery criteria assert it.
	Here the closed-world knowledge that licenses pruning is supplied by the closed shapes of a shape index, which let the engine soundly conclude that a region of the web cannot contribute to the query results.}

To perform pruning in \ac{LTQP} with shape indexes, an initial reachability criterion $C_0$ is defined.
This criterion must include a discovery reachability criterion $cd_{\text{shape index}}$ that leads to a shape index document.
After dereferencing a shape index $SI_i$, the query engine creates a set of links $IRI_p$ containing the links to prune.
The links to prune are identified by evaluating the shape index to find \acp{IRI} that are not relevant to the query, such that Equation~\ref{eq:evalQueryStructuralAssumption} holds, given that $G^{\prime}$ is produced using $C_0$.
This is done by performing a query-shape subsumption check ($\sqsubseteq_{qs}$), defined in the next section.

We define, given a query $Q$,
$$
	IRI_p = \left\{ \bigcup SI_i(s_j) \mid Q \not\sqsubseteq_{qs}  s_j \land s_j \in \text{dom}(SI_i) \right\}
$$
From these sets of links we define a pruning reachability criterion.
\begin{equation}
	cp_{si}(t, iri, Q) = iri \notin IRI_p
\end{equation}
The new reachability $C_i$ is created by taking the $Cd$ and $Cp$ of $C_{i-1}$ and adding $cp_{si}$ to $Cp$.

\rone{
	\paragraph{Example}
	In Figure~\ref{fig:dkg}, the engine's criterion $C$ combines a discovery criterion $cd_{\text{shape index}}$ that follows the links to each subweb's shape index (taken disjunctively) with a pruning criterion $cp_{si}$ (taken conjunctively).
	The shape index of Subweb~3 maps the post, user, and like shapes.
	The query subsumes the post shape but neither the user nor the like shape, so $IRI_p$ collects the user and like resources; $cp_{si}$ prunes exactly those, while the post resources, for which every $cp$ holds and at least one $cd$ holds, stay reachable.
	The same check prunes all of Subweb~1, whose like and user shapes are not subsumed, and in Subweb~2 every resource except the comments that reply to the posts.
}

\paragraph{Limitations}
This approach has three main limitations.
First, it assumes that data providers maintain up-to-date shape indexes; outdated indexes may lead to incomplete results.
A similar criticism could be leveraged against the method exploiting \ac{VoID} descriptions~\cite{Montoya2017}.
Second, if the query-shape subsumption check requires dereferencing all documents, it becomes ineffective and may slow down query execution.
Third, the approach does not consider cases where querying irrelevant documents could uncover relevant ones via additional reachability criteria.
Addressing this would require translating these criteria into queries, which is beyond the scope of this paper.

\subsection{Query-Shape Subsumption}\label{sec:containment}

To determine whether the contents of a resource conforming to a shape is query-relevant, we define a \emph{query-shape subsumption} problem denoted as $Q \sqsubseteq_{qs} S$, meaning $Q$ is subsumed by $S$.
A common approach for validating shapes over an \ac{RDF} graph is to translate shapes into SPARQL queries~\sepfootnote{sf:recursiveShape}~\cite{labragayo2017validatingdescribinglinkeddata, Corman2019, spapeExpressionConvert, delva2023}.
We denote the transformation of a shape $S$ into a query as $T(S)$, which yields a query $Q_s$.
We transform open shapes as queries over the entire \ac{KG}, since they impose only the minimal constraints required of a \ac{KG}.
When the problem is expressed as $\text{GSP} \sqsubseteq_{qs} Q_s$, we say that a \ac{GSP} is subsumed by $S$ if every result of the \ac{GSP} can be extended to a result of $Q_s$, and every triple pattern in a \ac{GSP} is equivalent to or a specialization of those in the $Q_s$.
Thus, the problem diverges from traditional query containment and query subsumption~\cite{Spasi2023, Pichler2014} under set semantics.
Query-shape subsumption does not consider only the set of solution mappings but also the constraints of the queries.
\footnote{In this sense, it bears some similarity to query containment under bag semantics, particularly through the notion of ``goals-onto'' containment mappings~\cite{Chaudhuri1993, Afrati2010}.}
The complexity of the problem is reduced by the fact that $Q_s$ has a \ac{GSP} structure where predicates are always \acp{IRI}.
This structure arises because shapes describe constraints on the predicate and object terms of sets of triples.
By exploiting this structure, it is possible to design an algorithm with polynomial-time complexity.
Moreover, empirical studies suggest that real-world queries tend to be relatively small~\cite{Doan2012, Bonifati2019}, making this algorithm applicable in practice.

More formally, for a $S$ to subsume a shape \ac{GSP}, we consider the queries $Q$ and $Q_s$, where $Q_s$ is the query translated from $S$, and both are of the form
$$
	Q_i = Q_{\text{body}} \bowtie Q_{\text{unions}}
$$
where $Q_{\text{body}}$ denote the \ac{BGP} of the query, and
$Q_{\text{unions}} = \bigcup Q_u$ represent the \acp{UGP}~\cite{w3SPARQLQuery} expressed in normal form, meaning
each $Q_u$ is of the form $q_i \cup q_j$, where each $q$ is a \ac{BGP} contain no union statements.

\rone{
	\paragraph{Example}
	The query in Figure~\ref{fig:dkg} has two \acp{GSP}.
	The \texttt{?message} star captures a subset of a post's predicates (\texttt{snvoc:hasCreator}, \texttt{snvoc:content}) and requires \texttt{rdf:type}~\texttt{snvoc:Post}; to decide $Q \sqsubseteq_{qs} S$ for the post shape of Subweb~3, that shape is translated into $Q_s = T(S)$, and every triple pattern of the \texttt{?message} star is equivalent to or a specialization of a pattern in $Q_s$ while each solution extends to one of $Q_s$, so the star is subsumed and the post resources are query-relevant.
	The \texttt{?linkedMessage} star captures a comment that must reply to a post (\texttt{snvoc:reply}~\texttt{?message}) and be of \texttt{rdf:type}~\texttt{snvoc:Comment}; it is subsumed analogously by the comment shape of Subweb~2.
}

\subsubsection{Algorithm}

\begin{algorithm*}[h]
   \centering
    \caption{Check if a \ac{GSP} is subsumed by $Q_s$ ($subsums_{\mathrm{graph star}}$)}\label{alg:containmentTree}
    \begin{algorithmic}[1]
       \scriptsize
 
       \REQUIRE  $Q_{star}$, $Q_{starG}$, $Q_s = Q_{s\mathrm{body}} \bowtie Q_{s\mathrm{unions}}$, $Eval_{star}$
       \ENSURE \TRUE $ $ or \FALSE $ $ whether the root of a \ac{GSP} $Q_{star}$ subsumes a shape.
 
       \IF{$M(S_{star}(Q_{star})) \in Eval_{star}$}
          \RETURN $M(S_{star}(Q_{star}))$
       \ENDIF 
 
       \FORALL{$tp \in Q_{star}$} 
          \IF{\NOT $match(tp, Q_{s\mathrm{body}})$}
             \STATE $hasOnePath \gets $ \FALSE
             \FORALL{$q_{us} \in Q_{s\mathrm{unions}}$}
                \IF{$subsums_{\mathrm{graph star}}(Q_{star}, Q_{starG}, q_{us}, Eval_{star})$}
                   \STATE $hasOnePath \gets $ \TRUE
                \ENDIF
             \ENDFOR
             \IF{\NOT $hasOnePath$}
             \STATE $Eval_{star} \gets Eval_{star} \cup (M(S_{star}(Q_{star})) \mapsto \mathrm{false})$
                \RETURN \FALSE
             \ENDIF
          \ELSE
             \IF{$O(tp) \in  S_{star}(Q_{starG})$}
               \STATE $Eval_{star} \gets Eval_{star} \cup (M(S_{star}(Q_{star})) \mapsto \mathrm{true})$
                \IF{\NOT $subsums_{\mathrm{graph star}}(Q_{star_{O(tp)}} \in Q_{starG}, Q_{starG}, Q_s, Eval_{star})$}
                  \STATE $Eval_{star} \gets Eval_{star} \cup (M(S_{star}(Q_{star})) \mapsto \mathrm{false})$
                   \RETURN \FALSE
                \ENDIF
             \ENDIF
          \ENDIF
       \ENDFOR
 
       \RETURN \TRUE
    \end{algorithmic}
 \end{algorithm*}

We define the function $subsums_{\text{graph star}}$ in Algorithm~\ref{alg:containmentTree} to evaluate whether a \ac{GSP} with a root star pattern $Q_{star_i}$ from $Q_{starG_i}$ is subsumed by $Q_s$.
The algorithm also takes a set $Eval_{star}$ to track which partial \acp{GSP} have already been evaluated.
The algorithm iterates over each triple pattern in the root star pattern $Q_{star_i}$ and uses the $match$ function to check if there exists a triple pattern in the \ac{BGP} of $Q_s$ whose domain of matched triples is a superset of that of the current pattern.
If the triple pattern cannot be found in the \ac{BGP}, the algorithm then looks into the \acp{UGP} of $Q_s$.
If an equivalent triple pattern is found, the algorithm checks whether the object of the triple pattern is the subject of a partial \ac{GSP} in $Q_{starG_i}$.
In this case, the algorithm recursively applies the same procedure to this partial \ac{GSP}.
To avoid cycles and redundant evaluations we maintain a set of evaluated answers in $Eval_{star}$.
To solve $Q \sqsubseteq_{qs} S$, we need to consider the number of \acp{GSP} from the \ac{BGP} with their number of segments in the \ac{UGP} and the number of \acp{BGP} in the \acp{UGP}.
This procedure is detailed in Algorithm~\ref{alg:containmentQuery}.
The following paragraphs analyze the time complexity of the algorithm.

\subsubsection{Time Complexity Analysis}

\paragraph{Worst-case per $Q_{star}$}
Let $n_{qstar}$ denote the number of distinct $Q_{star} \in Q_{starG_i}$.
For each node $Q_{star}$, the algorithm iterates over its triple patterns (line 4 to 25), resulting in a time complexity of $O(n_{tp_{Q_{star}}})$.
For each triple pattern that does not match the shape body ($Q_{s\mathrm{body}}$), the algorithm iterates over all union branches in $Q_{sunion}$ (line 7 to 11), making at most $n_{sunion}$ recursive calls.
The algorithm traverses the $q_{us}$ graph however each $q_{us} \in Q_{sunion}$ cannot contain a \ac{UGP}, and is therefore always of the form $Q_s = Q_{s\mathrm{body}}$, making this
branch (line 5 to 15) of the algorithm after a first execution not the worst case with a complexity of $O(n_{tp_{Q_{star}}}^{2} \cdot n_{sunion})$.

After the first execution, the worst-case scenario becomes one in which the triple patterns matches a pattern in $Q_{s\mathrm{body}}$, and the condition $O(tp) \in S_{star}(Q_{starG_i})$ holds (line 17 to 22).
In such cases, the algorithm recursively explores the corresponding partial \ac{GSP} by executing $subsums_{\mathrm{graph star}}$, following a graph traversal paradigm.

\paragraph{Tree Traversal Argument}
Although $Q_{starG_i}$ is a graph, the algorithm avoids cycles due to ``caching'', once a node $Q_{star}$ is evaluated, its result is stored in $Eval_{star}$.
Therefore, each node is visited at most once, and the overall traversal is equivalent to a tree traversal of size $n_{qstar} \leq |Q_{starG_i}|$.

\paragraph{Total Complexity}
Let $n_{tp_{Q_{star}}}$ denote the number of triple patterns in a particular $Q_{star}$, and let $n_{tp}$ be the total number of triple patterns across all nodes, so
$$
	n_{tp} = \sum_{Q_{star} \in Q_{starG_i}} n_{tp_{Q_{star}}}
$$

Then the total number of recursive operations over all nodes is bounded by:
$$
	O\left(\sum_{i=1}^{n_{qstar}} n_{sunion} \cdot n_{tp_{Q_{star_i}}}^{2}\right)
$$

$$
	O\left(n_{sunion} \cdot \sum_{i=1}^{n_{qstar}} n_{tp_{Q_{star_i}}}^{2}\right)
$$

$$
	O\left(n_{sunion} \cdot \left(n_{tp}^{2} - \sum_{i=1, j=1, i \neq j}^{n_{qstar}}  n_{tp_{Q_{star_i}}} \cdot  n_{tp_{Q_{star_j}}}\right)\right)
$$

$$
	O\left(n_{sunion} \cdot n_{tp}^{2}\right)
$$

\paragraph{Time Complexity of the $subsums_{\mathrm{Q}}$ Algorithm}

\begin{algorithm}[h]
    \caption{Check if a query $Q$ subsumes a $Q_s$ ($subsums_{\mathrm{Q}}$)}\label{alg:containmentQuery}
    \begin{algorithmic}
        \scriptsize

        \REQUIRE  $Q$, $Q_s$ and $Eval_{star}$
        \ENSURE \TRUE $ $ or \FALSE $ $ whether the shape subsumes $Q$.

        \FORALL{$Q_{starG_s} \in Q$}
            \FORALL{$Q_{star_s} \in Q_{starG_s}$}
                \IF{$subsums_{\mathrm{graph star}}(Q_{star_s}, Q_{starG_s}, \emptyset)$}
                    \RETURN \TRUE
                \ENDIF
            \ENDFOR
        \ENDFOR
        \RETURN \FALSE
    \end{algorithmic}
\end{algorithm}

The time complexity of the $subsums_{\mathrm{Q}}$ algorithm is straightforward to derive, as it consists of iterating over the \acp{GSP} of the query and applying the $subsums_{\mathrm{graph star}}$ algorithm. Thus, the time complexity is given by:
\begin{equation}
	O\left(n_{sunion} \cdot n_{tp_{max}}^{2} \cdot n_{Q_{star_{max}}} \cdot n_{Q_{starG}}\right)
\end{equation}

Where, $n_{tp_{max}}$ is the maximum number of triple patterns in any $Q_{starG}$,
$n_{Q_{star_{max}}}$ is the maximum number of $Q_{star}$ patterns in any $Q_{starG}$, and
$n_{Q_{starG}}$ is the number of \acp{GSP} in the query $Q$.\footnote{
	Sharing the evaluation results $Eval_{star}$ from $subsums_{\mathrm{graph star}}$ could reduce execution time and potentially the algorithm's complexity.
}

\section{Experimental Evaluation}\label{sec:experiment}

We implemented our approach using the \ac{LTQP} version of the Comunica query engine~\cite{taelman_iswc_resources_comunica_2018}.
We chose Comunica due to its modularity~\cite{taelman_swj_componentsjs_2022} and its established use in several \ac{LTQP} studies~\cite{Bogaerts2021LinkTW, Taelman2023, eschauzier_quweda_linkqueue_2023, Hanski2024, eschauzier_amw_rcubemetric_2024, tam2024opportunitiesshapebasedoptimizationlink}.
All implementations are open-source and are provided in the \hyperref[sec:supplementalMaterial]{supplementary material}.
Similarly to other \ac{LTQP} studies, we used SolidBench~\cite{Taelman2023}, which is based on the \ac{LDBC} social network benchmark~\cite{Angles2020}, to evaluate our contribution.
Furthermore, SolidBench was developed because no \ac{LTQP} benchmark existed prior to its creation~\cite{hartig2018linked, Taelman2023}, and, to the best of our knowledge, no other benchmarks have been introduced since.
We created an open-source module to generate shape indexes in SolidBench, based on user-provided mappings between \ac{ShEx} shapes and data model objects.
The shape-annotated portion of the data model includes posts, comments on posts, user profiles, user settings, varia data (unstructured data), cities, and likes.
The datasets are Solid Pods~\cite{sambra_solid_2016, dedecker2022s}.
In this paper, we consider a Solid Pod as a web-based file system that follows the \ac{LDP} specification~\cite{w3LinkedData}.
Each Solid Pod, alongside contains its data, a shape index and separate resources for each shape definition.
Some shapes are nested within others. 
For example, user profiles are associated with cities, and comments are associated with posts.
Depending on the pod instance, certain data model objects are materialized in a single file, while others are distributed across multiple files.
The benchmark provides queries that simulate typical read actions in social media use cases, such as retrieving replies to posts or identifying users connected to a given user. 
Queries may range from being highly selective with small result sets to broad with larger result sets, and in some cases can also be characterized as data-model selective.
The datasets contained approximately 4,200,000 triples and 1,528 subwebs, which, in this context, are Solid Pods.
The shape indexes contain 13 triples each, while the largest shapes have up to 150 triples. 
This can be considered insignificant, particularly because the number of triples does not scale with the size of the subwebs.
The entire data model and query templates are available in the \hyperref[sec:supplementalMaterial]{supplementary material}.
\rtwo{A description of each query template used in our evaluation is provided in Table~\ref{tab:queryTemplateDescription} in the \hyperref[sec:appendix]{appendix}.}

To evaluate our approach, we conducted the following experiments: 
\begin{itemize}
   \item We measured the execution time and results of our query-shape subsumption algorithm using the shapes from the study.
   \item We then compared our shape index approach to state-of-the-art Solid Pod network traversal algorithms: one leveraging the type index specification~\cite{Taelman2023}, and another using the \ac{LDP} specification~\cite{Taelman2023}, in a network where each Solid Pod provides a complete shape index.~\footnote{It should be noted that the type index and shape index approach are extensions of the \ac{LDP} approach.}
   \item We analyze the relationship between the number of \ac{HTTP} requests and query execution time, drawing on data collected across all experiments.
\end{itemize}

Subsequently, we assessed the resilience of our approach by gradually reducing the shape index information across the network by performing the following experiment:

\begin{itemize}
   \item We measured query execution time in networks where 0\%, 20\%, 50\%, and 80\% of Solid Pods expose a shape index.
   \item We measured query execution time in networks where 20\%, 50\%, and 80\% shape indexes are complete.
   \item We measured query execution time in networks having shapes that incorporate only data from the Solid Pods and shapes providing a minimal dataset description where the object constraints are always an \ac{IRI} or a literal.
\end{itemize}

We conducted the experiment using queries from five different instantiations of SolidBench query templates, varying the starting pods in a random yet reproducible manner.
Experiments were repeated 50 times with a 2 minute timeout per query execution. 
They were conducted on an Ubuntu 20.04.6 LTS machine with a 2x Hexacore Intel E5645 CPU and 24GB RAM.


\section{Results and Discussion}\label{sec:result}
\subsection{Evaluation Against Other Approaches}

\begin{figure*}
	\centering
	\includegraphics[width=\linewidth]{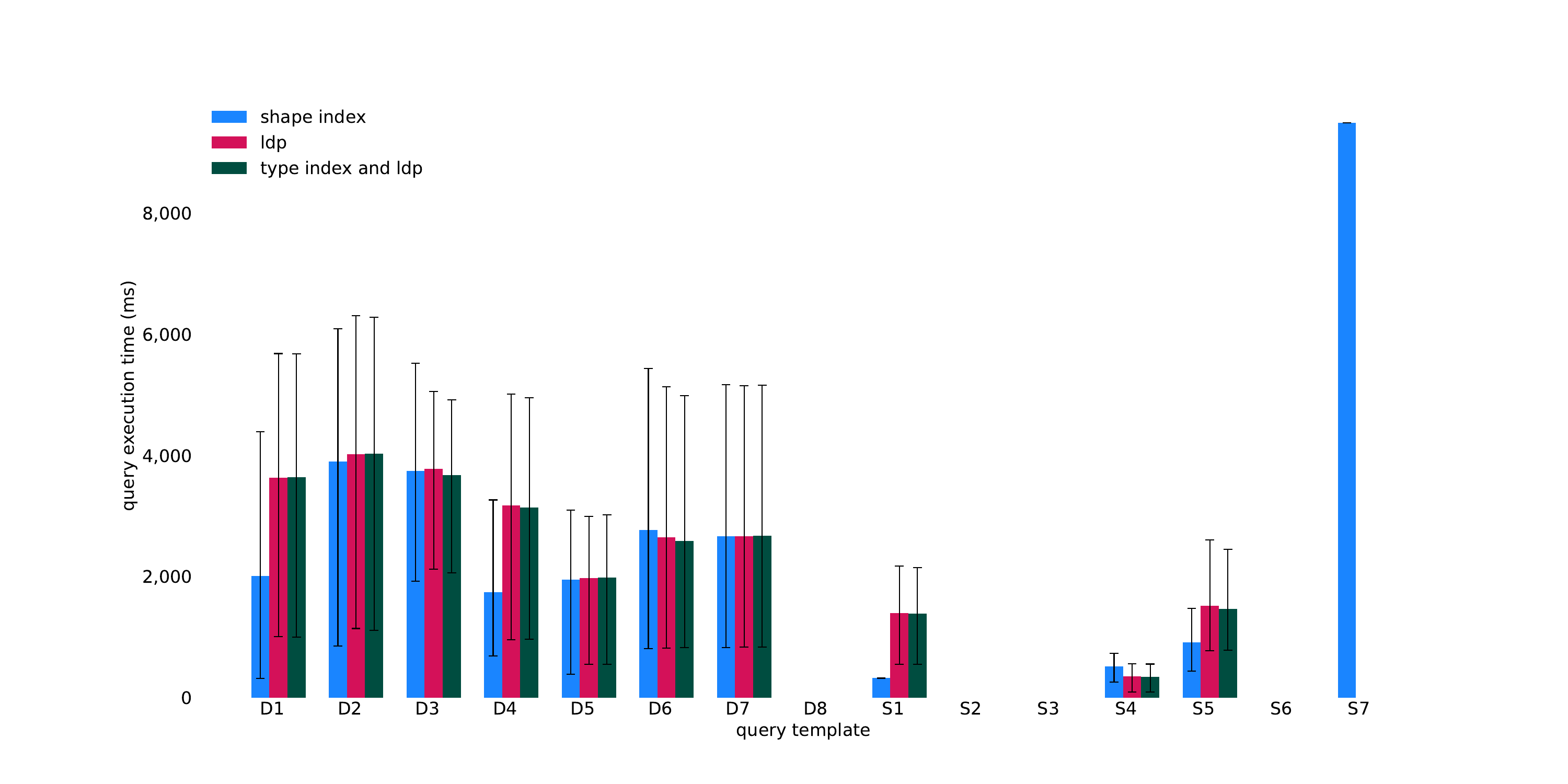}
	\caption{
		Performance comparison of different methods for query execution.
		The shape index performs better than both the type index and \ac{LDP} for data-model selective queries.
		The shape index is also able to answer a query from template S7 which was not possible with the other approaches.
		For non-data-model selective queries, execution time differences are not statistically significant, as shown in Table~\ref{tab:statSignificanceStateOfTheArt}.
		An exception occurs with queries of template S4, which do not leverage the structure of the publication environment.
		Results are presented by query template, leading to a multimodal distribution; therefore, variance analysis does not indicate statistical significance.
	}
	\label{fig:compApproach}
\end{figure*}

\begin{figure}
	\def\svgwidth{\linewidth}
	\input{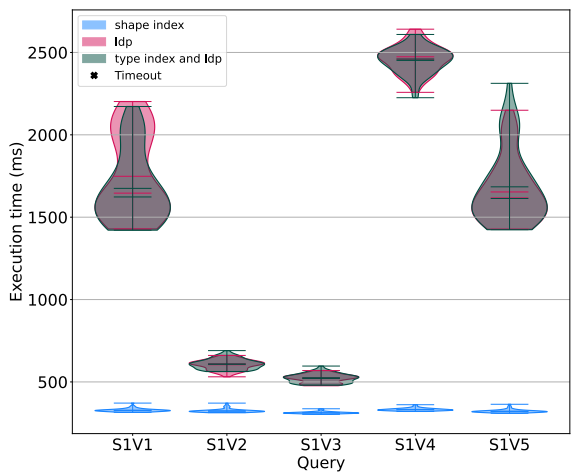}
	\caption{
		Distribution of query execution times (sample size: 50) for queries derived from the S1 template.
		The results show that the execution time of data-model selective queries is significantly reduced when using the shape index approach.
	}
	\label{fig:violonS1}
\end{figure}

\begin{figure}
	\def\svgwidth{\linewidth}
	\input{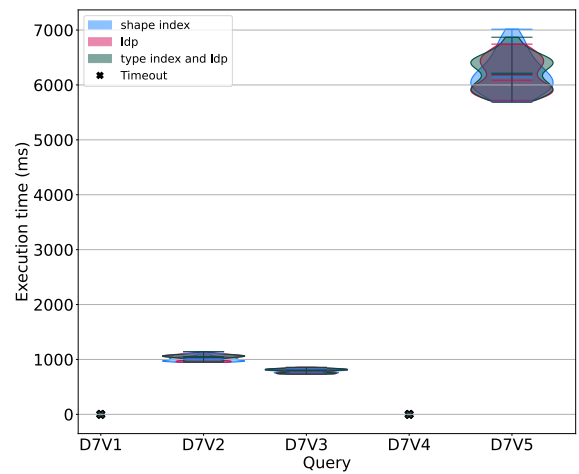}
	\caption{
		Distribution of query execution times (sample size: 50) for queries derived from the D7 template.
		The results show that the execution time of non-data-model selective queries is not significantly affected by the shape index approach.
	}
	\label{fig:violonD7}
\end{figure}

\begin{figure}
	\def\svgwidth{\linewidth}
	\input{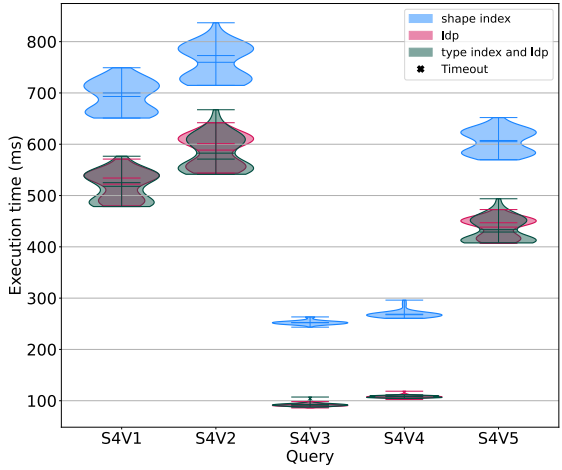}
	\caption{
		Distribution of query execution times (sample size: 50) for queries derived from the S4 template.
		Queries that do not exploit the structural properties of the dataset are significantly negatively impacted by the shape index approach.
	}
	\label{fig:violonS4}
\end{figure}

\subsubsection{Analysis of the Query Execution Time}
Figure~\ref{fig:compApproach} demonstrates that the shape index approach performs queries faster than or comparable to state-of-the-art Solid Pod network traversal algorithms across all query templates except S4.
In the \hyperref[sec:appendix]{appendix} Figure~\ref{fig:ratioApproaches} presents the ratio of execution time between the type index approach and the shape index approach.
Table~\ref{tab:statSignificanceStateOfTheArt} presents the statistical significance by query templates.
The shape index approach allows for successful termination of S7 template queries, which other approaches cannot complete due to timeouts resulting from an excessive number of \ac{HTTP} requests.
Query processing with the shape index approach requires as little as 13\% of the execution time needed by the type index approach, achieving up to 7-fold performance improvement for S1 queries.
The queries that perform the best are those in which the number of \ac{HTTP} requests decreased the most, as can be inferred from the analysis of Table~\ref{tab:ratioUsefulResources}.
This table presents the average percentage of query-relevant resources per query template, derived from the where-provenance~\cite{buneman2001and} of the query results and the number of \ac{HTTP} requests.
Queries with a high decrease of the number of \ac{HTTP} request are known to be data-model selective due to the query-shape subsumption algorithm.
In our context, a query that requests a single data model object is classified as data-model-selective, since in our benchmark the subwebs contain at most five distinct data model objects.
From the analysis of the queries, those derived from templates D1, D4, S1, S3, and S7 (S3 is a template that cannot complete the execution before the timeout, while S7 can only be executed successfully with the shape index) are identified as data-model-selective and, as expected, exhibit a reduction in both the number of \ac{HTTP} requests and the query execution time.
Figure~\ref{fig:violonS1} shows the distribution of query execution times for each instance of the S1 query template.
The distributions for all query templates are provided in the \hyperref[sec:supplementalMaterial]{supplementary material}.
Queries from templates D6 and D7 show no reduction because they require nearly every data model objects in the datasets to be processed by the engine, making our approach ineffective in these cases.
\rtwo{The D6 and D7 templates are presented in Listings~\ref{lst:d6} and~\ref{lst:d7}.}
Figure~\ref{fig:violonD7} shows the distribution of query execution times of queries of the D7 templates.
We notice that queries from template S4 with the shape index performed worse in every instance, with an increase in query execution time of up to 2.80 times as shown in Figure~\ref{fig:violonS4}.
\rtwo{The S4 template is presented in Listing~\ref{lst:s4}.}
This is further illustrated in Table~\ref{tab:ratioUsefulResources}, which shows that for these queries, the type index traversal algorithm achieves a ratio of query-relevant resources dereferenced of 100\% or 50\%, compared to only 6\% with the shape index approach.
The poor performance is due to the fact that the links acquired by the other approaches were selected based on reachability criteria that did not leverage the structural properties of the dataset, such as in the case of \texttt{Cmatch}~\cite{hartig2016walking}, a reachability criterion based on the structure of the query.
In contrast, the shape index approach always enforces the use of these properties, resulting in additional \ac{HTTP} requests and increased processing time.
However, those queries were already efficient, with the type index traversal approach completing in only about 0.30\% of the maximum allowed execution time.
Nonetheless, these results still highlight a category of queries and networks for which our approach is not well-suited.
Those results mostly validate \textbf{H1} and \textbf{H2} however the shape index approach can drastically increase the execution time when the structural properties are not used.

\renewcommand{\arraystretch}{1.2}
\begin{table*}
	\begin{center}
		\resizebox{\linewidth}{!}{\begin{tabular}{|c|c|c|c|c|c|c|c|c|c|c|}
			\hline
            Approach/ Query Template & D1 & D2 & D3 & D4 & D5 & D6 & D7 & S1 & S4 & S5 \\
			\hline
            shape index (\%) & $\boldsymbol{14^{42}_{6}}$ & $\boldsymbol{38^{91}_{11}}$ & $\boldsymbol{52^{62}_{37}}$ & $\boldsymbol{15^{20}_{3}}$ & $\boldsymbol{9^{13}_{2}}$ & $18^{41}_{7}$ & $8^{11}_{2}$ & $\boldsymbol{12^{12}_{12}}$ & $6^{6}_{6}$ & $\boldsymbol{10^{11}_{4}}$ \\
            \hline
			type index (\%) & $4^{6}_{4}$ & $35^{86}_{8}$ & $51^{61}_{36}$ & $9^{15}_{3}$ & $8^{12}_{2}$ & $\boldsymbol{20^{42}_{8}}$ & $\boldsymbol{9^{13}_{2}}$ & $4^{8}_{1}$ & $\boldsymbol{80^{100}_{50}}$ & $3^{4}_{2}$ \\
			\hline
		\end{tabular}}
	\end{center}
	\caption{
		\textbf{The percentage of query-relevant resources.}
        For most queries, the percentage of query-relevant resources is low (cell values are shown as $\text{avg}^{\text{max}}_{\text{min}}$). 
		Shape index generally matches or outperforms the type index, except for templates D6, D7, and S4.
		The difference for S4 is more pronounced, reaching 100\% with the Type Index compared to only 6\% with the shape index.
		}
	\label{tab:ratioUsefulResources}
\end{table*}

\subsubsection{Temporal Analysis}
To further our analysis, we compare three temporal metrics: the arrival time of the first results, the termination time, defined as the duration between the arrival of the last result and the end of query execution, and the waiting time, which we define as the accumulated time gaps exceeding one second between the reception of consecutive results.
We choose a 1 second threshold for waiting time based on research in responsive system design.
These waiting time guidelines are derived from the field of neuropsychology and are considered stable regardless of technological expectation for human users~\cite{uxtigersNeedSpeed, Nielsen1993}.
A waiting time below 0.1 seconds is perceived as instantaneous, a delay of up to 1 second preserves a seamless flow of thought, while delays exceeding 10 seconds tend to cause user attention to drift~\cite{Nielsen1993}.
We additionally computed the diefficiency metric in relation to time (\textit{dief@t})~\cite{Acosta2017}, at the previously mentioned time intervals: 0.1\,s, 1\,s, and the arrival time of the final result.
We did not choose 10\,s as the threshold, because queries that terminate complete in under 10\,s.
Tables~\ref{tab:dief} and \ref{tab:continuousPerf} in the \hyperref[sec:appendix]{Appendix} present the average results for the metrics presented above.

\begin{figure}
	\centering
	\def\svgwidth{\linewidth}
	\input{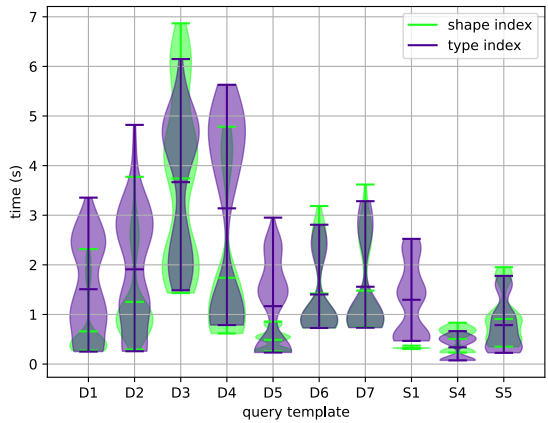}
	\caption{
		\textbf{The first result distribution by query template.}
		The shape index approach tends to produce the first results earlier than the type index approach, particularly for queries where the shape index achieves shorter total execution times.}
	\label{fig:first_res}
\end{figure}

\paragraph{For Most Queries, the Shape Index Approach Tends to Produce the First Results Faster}
Figure~\ref{fig:first_res} presents the time of arrival of the first results for all query instances, grouped by query templates.
The plot indicates that, for most queries, the shape index approach tends to produce results faster, as shown by distributions concentrated around lower first result arrival times.
Templates D1, D2, D5, and S1 exhibit the most significant improvements in first-result latency when using the shape index.
Surprisingly, although queries from templates D2 and D5 sometimes performed worse with the shape index approach, as shown in Figure~\ref{fig:compApproach}, they still produced the first result faster than with the type index approach.
This suggests that, although the shape index does not explicitly prioritize early result generation, the pruning it performs may implicitly favor faster first results arrival with some queries and networks.
This implicit prioritization does not apply uniformly across all templates.
For example, queries from template D3 exhibit faster execution times across most instances when using the shape index.
However, as shown in Figure \ref{fig:first_res}, the shape index also results in a higher mean and a longer tail for the time to first result.
This difference arises from the structure of the subwebs targeted by D3 queries.
In some cases, a result can be obtained from a document located one link away from the seed document.
The direct dereferencing using the type index approach can exploit this proximity, enabling early result generation.
In contrast, the shape index approach may delay first result production because it relies on discovering the shape index documents before exploring the rest of the subweb.

\begin{figure}
	\centering
	\def\svgwidth{\linewidth}
	\input{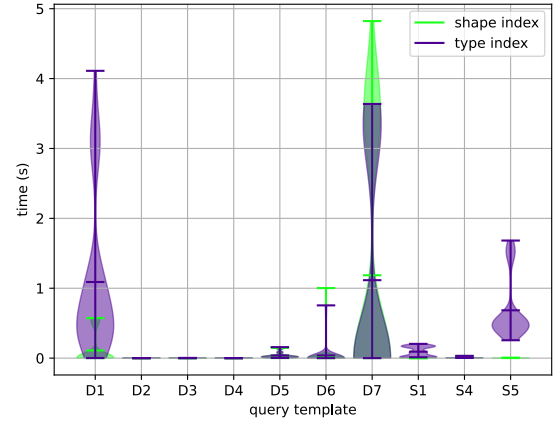}
	\caption{
		\textbf{The termination time distribution by query template.}
		With the exception of queries from templates D6 and D7, which have similar execution times, the shape index either maintains or improves query termination time.}
	\label{fig:termination_time}
\end{figure}

\paragraph{The Shape Index Approach Can Reduce the Termination Time}
Figure~\ref{fig:termination_time} presents the termination times for the query execution instances, grouped by template.
For queries where execution time improved and termination time was non-zero, such as those from templates D1, S1, and S5, the shape index approach significantly reduces termination time.
This improvement aligns with the substantial increase in the percentage of query-relevant resources dereferenced when using the shape index, as shown in Table~\ref{tab:ratioUsefulResources}.
A long termination time generally indicates that, toward the end of execution, a large number of non-query-relevant resources are being dereferenced.
Although Table~\ref{tab:ratioUsefulResources} does not show the temporal progression of this percentage it still give us information about the quantity of non-query-relevant resources being dereferenced.
For templates D5 and D6, the termination-time distributions for the shape index approach have a similar shape but exhibit a longer tail, particularly D6, which can take up to one second longer to terminate.
In these cases, Table~\ref{tab:ratioUsefulResources} shows that the proportion of query-relevant resources dereferenced using the shape index is within $\pm$1\% of that obtained with the type index.
A similar situation is observed for template D3, although in this case the type index achieves a termination time of zero.
The longer termination times for D5 and D6 may be attributed to the additional dereferencing of shape index documents.
Notably, for template D6 in Figure~\ref{fig:compApproach}, an outlier query increases execution time by approximately a ratio of 1.2, which could explain the extended tail and the additional second of termination time.

\begin{figure}
	\centering
	\def\svgwidth{\linewidth}
	\input{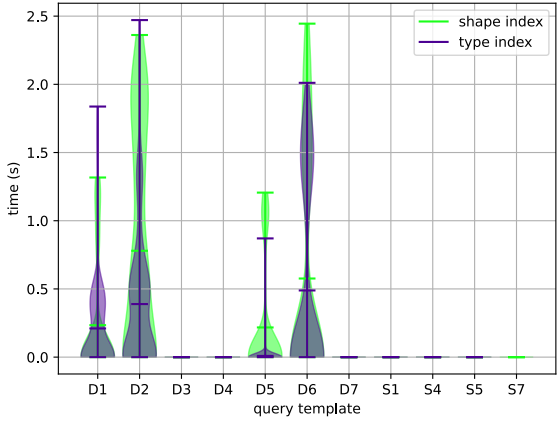}
	\caption{
		\textbf{The waiting time distribution by query template.}
		Except for queries from template D1, the shape index approach tends to either maintain or worsen the waiting time.}
	\label{fig:waiting_time}
\end{figure}

\paragraph{The Shape Index Approach Generally Increases the Waiting Time}
Figure~\ref{fig:waiting_time} presents the waiting time for the query instances, grouped by template.
With the exception of queries from the D1 template, which show a decrease in waiting time, the shape index approach generally increases the waiting time for queries that experience a delay before producing results.
This increase is likely due to the additional dereferencing of shape index documents that must occur before relevant results can be generated.
The shape index must also be retrieved whenever traversal reaches a new subweb, which can stall the engine between two consecutive results and widen the gaps counted as waiting time, even when the total query execution time is reduced or remains unchanged.

\begin{figure}
	\centering
	\def\svgwidth{\linewidth}
	\input{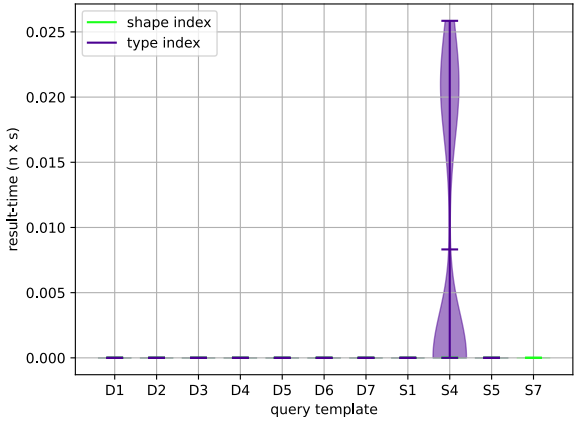}
	\caption{
		\textbf{The \textit{dief@0.1s} distribution by query template.}
		At 0.1 second, the value is zero for every query template except S4. This is expected, as for all other templates the first result arrives after 0.1 seconds.}
	\label{fig:dief_01}
\end{figure}

\begin{figure}
	\centering
	\def\svgwidth{\linewidth}
	\input{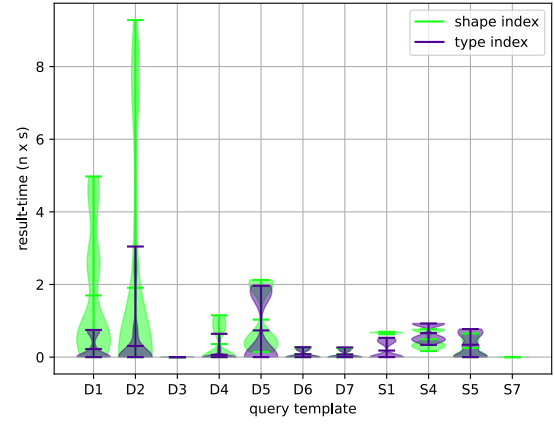}
	\caption{
		\textbf{The \textit{dief@1s} distribution by query template.}
		At 1 second, the shape index approach shows a higher \textit{dief@1s} (\textbf{higher is better}), indicating that more results can be obtained by users while maintaining a seamless flow of thought.}
	\label{fig:dief_1}
\end{figure}

\begin{figure}
	\centering
	\def\svgwidth{\linewidth}
	\input{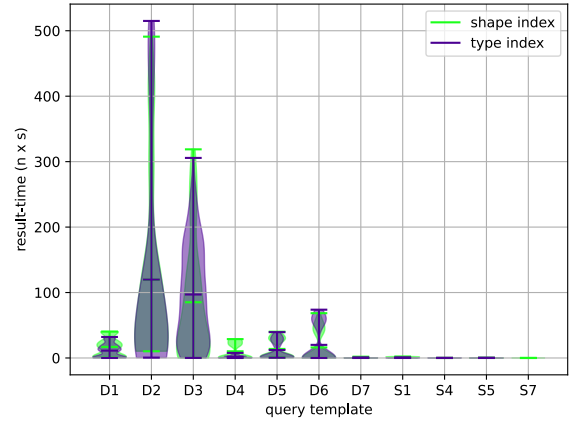}
	\caption{
		\textbf{The \textit{dief@lr} distribution by query template.}
		At the time of the last result, the shape index approach tends to exhibit a similar \textit{dief@t} (\textbf{higher is better}).}
	\label{fig:dief_lr}
\end{figure}

\paragraph{The Shape Index Approach Can Improve the Diefficiency Metric}
Figures~\ref{fig:dief_01}, \ref{fig:dief_1}, and~\ref{fig:dief_lr} show the \textit{dief@t} values measured at 0.1 seconds (\textit{dief@0.1s}), 1 second (\textit{dief@1s}), and at the time of the last result (\textit{dief@lr}).\footnote{\label{fn:dief_lt}This means that if, for a given query, the shape index approach produces the last result at 5~seconds and the type index at 5.10~seconds, then the \textit{dief@t} metric will be computed at 5.10~seconds.}
The \textit{dief@0.1s} results show that, except for S4, no approach is able to produce instantaneous results, which aligns with the observations in Figure~\ref{fig:first_res}.
Overall, this indicates that, with the current approaches and benchmark, producing instantaneous results is not feasible.
The \textit{dief@1s} results indicate that the shape index approach generally produces a higher number of early results, except for queries from templates S4 and S5.
This result is expected for S4 as the performance with the shape index is drastically reduced as shown in Figure~\ref{fig:compApproach} however with S5 it is less expected as it is a query template that perform better with the shape index.
It would be expected to see a performance improvement in terms of \textit{dief@t} at the last results for S5, however, the values are zero or close to zero.
This is because these queries produce only a single result, so \textit{dief@t} naturally tends toward 0 unless there is a very large difference in execution time between the two approaches.~\ref{fn:dief_lt}
These results suggest that, for fast early result arrival, the shape index approach tends to performs better.
This finding is particularly relevant in the context of social media applications, where it is often more important to obtain fast, relevant results to maintain interactivity than to retrieve complete results.
The \textit{dief@lr} show a similar behavior but with lesser performance gain compared to \textit{dief@1s} (see Table~\ref{tab:dief} to compare the average values).


\subsection{Query-Shape Subsumption Evaluation} \label{sec:experimentAlgoSubsumption}

The empirical evaluation of the query-shape subsumption algorithm shows that its execution time with the more detailed shapes from our experiment is negligible, with a maximum execution time of 5ms (0.0039\% of the timeout).
Table~\ref{tab:queryShapeContainmentEval} present these results.
This outcome is expected, as the algorithm has polynomial time complexity, and the shapes and queries in the experiments are small and not deeply nested.
This result validates \textbf{H3}.

\subsection{Evaluation of the Resilience of the approach}

\begin{figure*}
	\centering
	\includegraphics[width=1\linewidth]{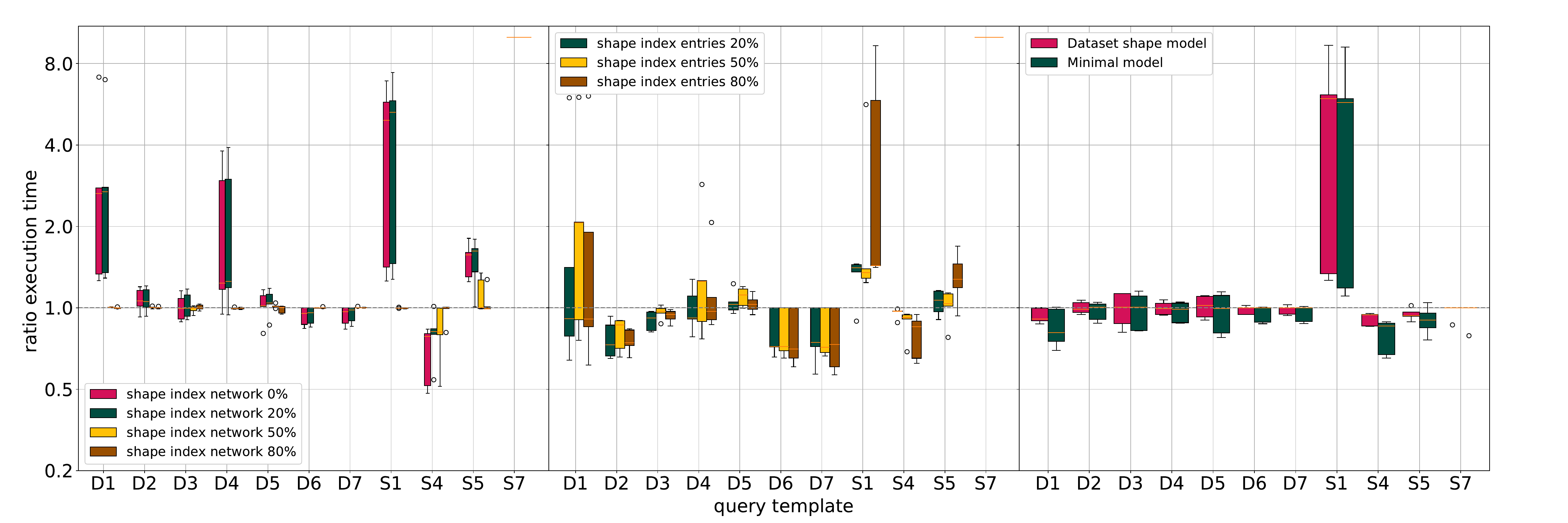}
	\caption{
		Shape index approaches tend to perform less effectively with limited network information and comparatively better where the baseline shape index underperforms.
		A higher ratio indicates a longer query execution time compared to a network with complete shape index information (\textbf{lower is better}).
	}
	\label{fig:adaptShapeIndex}
\end{figure*}

The final part of the results analysis focuses on the resilience of the shape index approach.
In this analysis, we examine the impact of reducing the shape index information in the network and compare the results with a network in which all pods are exposed to detailed, complete shape indexes.
Figure~\ref{fig:adaptShapeIndex} presents three plots that illustrate the results of our evaluation of the approach's resilience.
The plot on the left shows the variation in the availability of shape indexes across the network.
As expected, we observe that queries that performed better in Figure~\ref{fig:compApproach} tend to perform worse with reduced shape index information, while queries that performed poorly improved.

The plot in the middle shows the variation in the percentage of shape index entries using closed shapes.
The results here are more nuanced.
While there is a general trend for query evaluations with a lower percentage of closed shapes to behave similarly to the plot on the left, we also observe both performance gains for some query templates and a drastic performance loss for the queries of S1 when 80\% of the shape entries are closed.
The performance gain occurs because not every entry needs to be closed to prune the query irrelevant documents.
Entries mapped to an open shape are always considered relevant because the shape translates to a query fetching the whole \ac{KG}.
If the subsumption check leads to the same conclusion, then, given a closed entry, the execution will be more expensive because
when shapes are nested, the nested shapes need to be dereferenced to solve the query-shape subsumption algorithm.
For the queries of S1, with 80\% of closed shape entries, the performance lost was due to random chance, as the discriminatory entries were provided with open shapes in multiple instances when looking at the raw data.

The right plot shows the variation in the level of detail of the shapes by reducing their detail.
Since the shapes are closed in this experiment the level of detail was varied by changing the constraint of the object terms.
Most queries tend to perform similarly or better, except for those of S1.
Upon analyzing the output of our query subsumption algorithm, we observe that the additional information to the shape provided in our base approach does not affect the algorithm's results.
However, in the baseline shape index experiment, the engine must dereference more shapes, potentially increasing execution time.
Queries of template S1 are the only ones where the added information can discriminate multiple parts of the datasets' domain leading to worse performance.
This indicates that in some situations, adding more information can be beneficial.
This sensitivity of the quantity of the information in the index also helps to explain the results for S1 queries in the middle plot.
In that case, the engine still had to dereference shapes from each dataset, and the information available was likely insufficient to significantly discriminate between resources.

These results show that \textbf{H4} and \textbf{H5} are rejected: reducing information is not always detrimental to query execution.
It becomes detrimental only when the omitted information is critical for determining the query relevance of resources.
Therefore, the outcome is dependent on both the query and the network.
Moreover, the completeness of shape indexes has a smaller impact on performance than their presence or absence.
\textbf{H6} is accepted; it is possible to retain performance gains even in networks with reduced shape-index information.

\begin{figure*}[htbp]
	\centering
	\begin{minipage}[t]{0.45\linewidth}
		\centering
		\includegraphics[width=\linewidth]{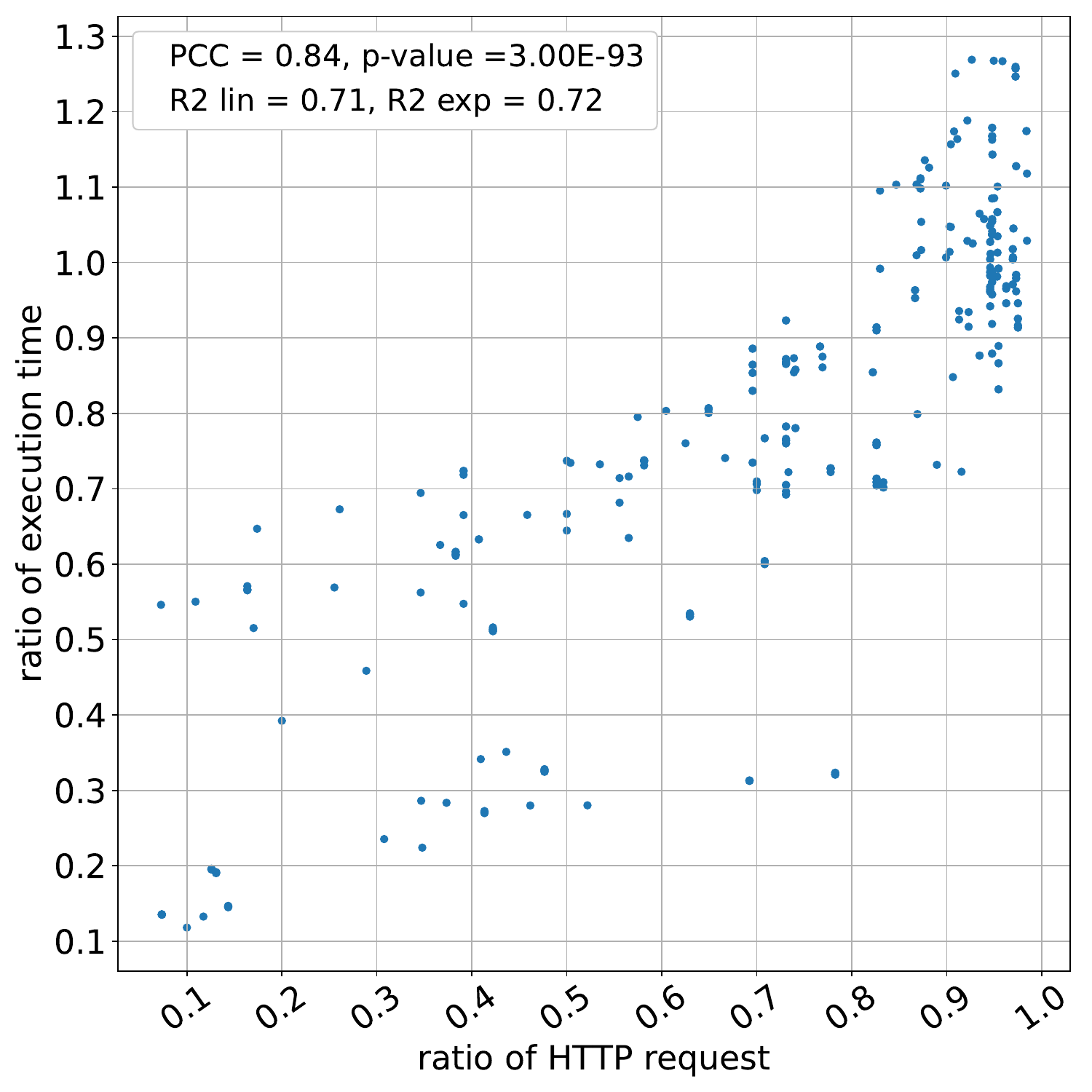}
		\label{fig:http_req_exec_time_cor_better}
	\end{minipage}
	\hspace{0.05\textwidth}
	\begin{minipage}[t]{0.45\linewidth}
		\centering
		\includegraphics[width=\linewidth]{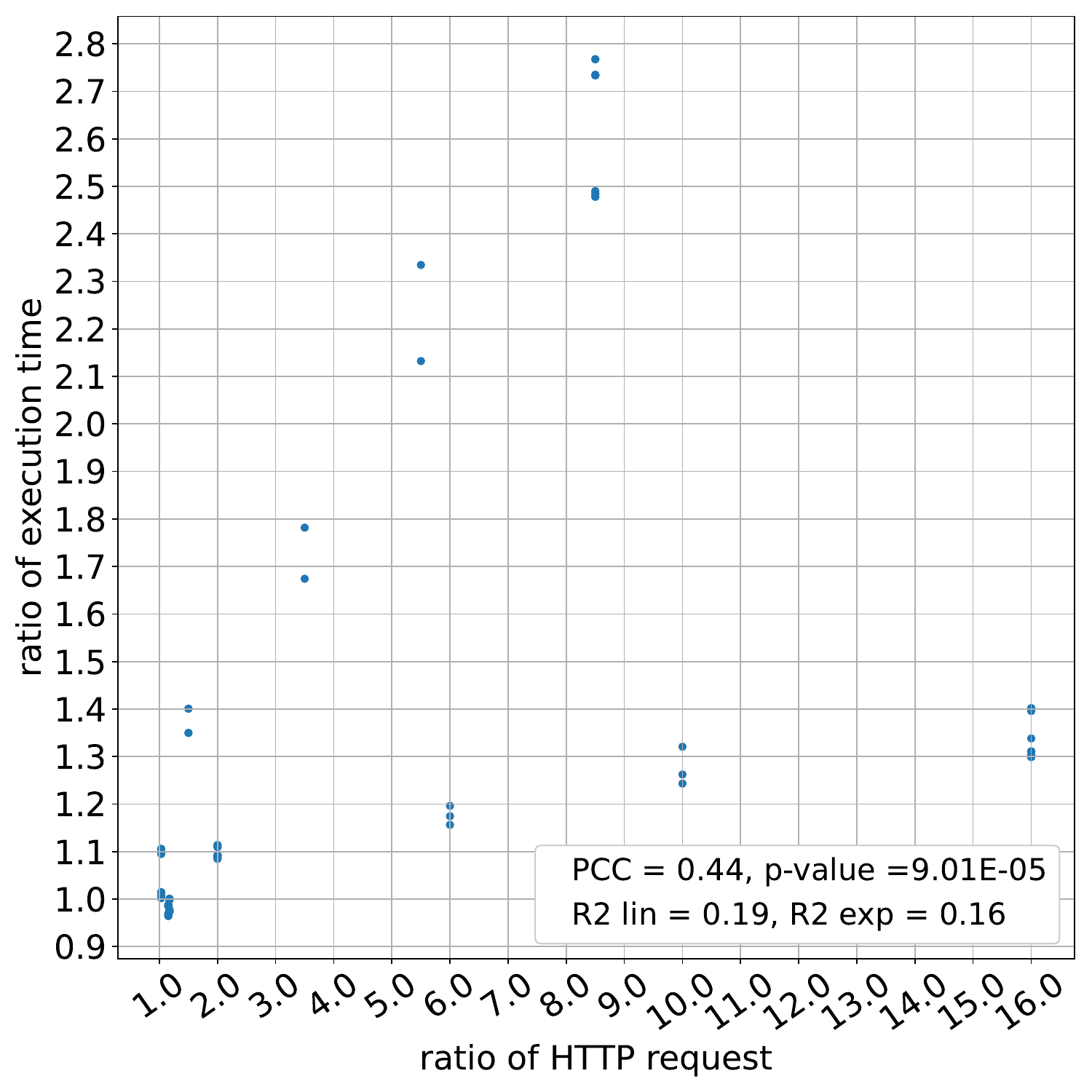}
		\label{fig:http_req_exec_time_cor_worse}
	\end{minipage}

	\caption{
		The data show two regimes in the relation between the number of \ac{HTTP} requests and the execution time.
		The correlation is more linear in the left figure than in the right figure.
	}
	\label{fig:http_req_exec_time_cor}
\end{figure*}

\subsection{Relationship Between HTTP Request and Query Execution Time}

To study the relationship between the number of \ac{HTTP} request and the query execution time, the ratio of \ac{HTTP} request of the different approaches and the ratio of query execution time was
calculated in relation to the type index approach.
Figure~\ref{fig:http_req_exec_time_cor} presents our analysis.
The relationship between \ac{HTTP} request and query execution time can be divided into two regimes.
In the first regime (left figure), where the shape index approach reduces the number of \ac{HTTP} requests, we notice a positive linear correlation with a \ac{PCC} of 0.84 and a high statistical significance ($< 0.01$).
We can notice that toward the end, the curve appears to exhibit a more exponential behavior.
Evaluating an $R^2$ score with an exponential best fit curve we get a score of 0.72 and 0.71 for a linear curve.
Above a ratio of approximately 0.85 of \ac{HTTP} requests, the shape index approach did not guarantee a reduction in query execution time.
There can be multiple explanations for this behavior.
First, the methods have some overhead due to the query-shape subsumption algorithm and the state retention of the pruning reachability criteria.
However, the query-shape subsumption algorithm execution time is negligible for one execution and the state retention has a polynomial time complexity, thus it should not have a high impact on the execution time.
Another explanation is the number of \ac{HTTP} requests that are performed in parallel.
The \ac{LTQP} version of Comunica performs 10 \ac{HTTP} requests in parallel, thus, we would expect that with a low number and ratio of \ac{HTTP} requests, the performance would remain largely unchanged or slightly worse, which is what can be observed in the curve.

In the second regime (right figure), the shape index increases the number of \ac{HTTP} requests.
We notice a moderate positive linear correlation with a \ac{PCC} of 0.44.
The overall correlation between reducing \ac{HTTP} requests and query execution time is positively linear, with a moderate linear correlation with a \ac{PCC} of 0.56 and a high statistical significance.
The correlation is more linear than exponential with $R^2$ scores respectively of 0.31 and 0.24, however due to the low score it is difficult to determine the nature of the distribution.

Explaining the two behavioral regimes observed in the data is challenging.
One possible explanation for the poorer performance of the shape index approach in certain cases could be the low number of samples.
However, we also observe that the relationship between the two variables differs significantly across the regimes.
In the first regime, the relationship is close to one-to-one, with a slope of approximately 0.91.
In contrast, the second regime shows a much flatter slope of around 0.08, suggesting that the ratio of \ac{HTTP} requests has a weaker influence.
This disparity indicates that the explanation may be more complex than just a sample size issue.
The increased number of \ac{HTTP} requests can be attributed to queries executed using the D6, D7, and S4 templates, with the S4 template causing the largest increase.
Queries from the S4 template typically require only 1 or 2 \ac{HTTP} requests, so when running with 10 concurrent requests, the impact on performance is limited.
On the other hand, queries using the D6 and D7 templates, executed with the type index traversal algorithm, required 26, 23, and 129 \ac{HTTP} requests.
This significantly higher number of \ac{HTTP} requests explains why the increase in \ac{HTTP} requests had a greater impact on query execution time.

These results partially validate \textbf{H7}, as a linear relationship is observed when there are a sufficient number of \ac{HTTP} requests and a significant reduction ratio, taking concurrent requests into account.

\subsection{\rone{Summary of Findings}}

\rone{Across the evaluation, the shape index approach reduced both the number of retrieved sources and the query execution time for data-model selective queries while preserving result completeness, but it degraded performance for queries that do not exploit the structural properties of the network, such as those of template S4.
	The query-shape subsumption algorithm introduced negligible overhead.
	Reducing the amount of shape index information was not consistently detrimental: neither stricter constraints nor more complete indexes guaranteed better performance, yet the approach retained its gains even under partial coverage.
	Finally, the reduction in \ac{HTTP} requests correlated with the decrease in execution time only when a sufficient number of requests were involved.
	Table~\ref{tab:hypothesisSummary} summarizes, for each hypothesis, whether the evaluation supports it.}

\begin{table*}[htbp]
	\centering
	\ifreview\color{revone}\fi
	\begin{tabular}{|c|l|c|}
		\hline
		\textbf{Hypothesis} & \textbf{Summary}                                                    & \textbf{Outcome}    \\
		\hline
		H1                  & Reduces sources and execution time for data-model selective queries & Supported           \\
		H2                  & No effect on non-data-model selective queries                       & Partially supported \\
		H3                  & Negligible query-shape subsumption overhead                         & Supported           \\
		H4                  & Stricter shape constraints reduce \ac{HTTP} requests more           & Rejected            \\
		H5                  & More complete shape indexes give faster execution                   & Rejected            \\
		H6                  & Gains hold under partial shape-index coverage                       & Supported           \\
		H7                  & Linear \ac{HTTP}-request / execution-time relationship              & Partially supported \\
		\hline
	\end{tabular}
	\caption{\rone{Summary of the hypotheses and whether the evaluation supports them.}}
	\label{tab:hypothesisSummary}
\end{table*}

\section{Conclusion}\label{sec:conclusion}

In this article, we introduced a pruning mechanism for \ac{LTQP}, extending the concept of reachability criteria and leveraging shape indexes.
Using the SolidBench benchmark, we demonstrate that our approach significantly reduces the number of \ac{HTTP} requests without degrading performance, particularly when queries exploit the structural properties of the dataset.
In the best-case scenario, query execution time improved by up to 7 times, and a previously unexecutable query completed successfully.
On average, performance improved by a factor of 1.76.
In the worst case, execution time increased by 2.8 times for a query that did not exploit the network structure; however, this query was already highly efficient, consuming only 0.30\% of the maximum allowed execution time.
Our method also handles scenarios with partial or reduced shape index information in networks, making it relevant for decentralization efforts that allow third-party clients to efficiently query large, heterogeneous datasets.

The approach imposes minimal overhead on servers, requiring only the serving of small static shape index files.
\rone{The construction and maintenance of shape indexes were beyond the scope of this work.
They are lightweight metadata that is low-cost to produce and maintain, as descriptive shape indexes can be generated automatically~\cite{fernandez2023extracting} and their \ac{URI}-template entries avoid exhaustive redefinition.}
This approach has the potential to benefit linked data-driven decentralization efforts such as the Solid ecosystem and support standardization efforts such as the Linked Web Storage Working Group~\footnote{\url{https://www.w3.org/groups/wg/lws/}} by improving query processing throughout these networks without requiring the maintenance of large centralized indexes between data providers.
This work opens a new line of research in integrating data models directly into decentralized query execution, particularly for non-indexed networks. 
Future work could focus on supporting filter expressions in the query-subsumption algorithm, enhancing query planning in \ac{LTQP}~\cite{taelman2024towards} with \ac{RDF} data shapes,
maintaining and developing shape indexes, and exploring other low-cost, low-maintenance indexing structures.

\paragraph*{Supplemental Material Statement:}\label{sec:supplementalMaterial} All source code, benchmark queries, datasets, raw results, and supplementary analyses are available online.~\footnote{
    \ifanonymous
       \url{https://anonymous.4open.science/r/documentation-1A65}
    \else
       \url{https://github.com/shapeIndexComunicaExperiment/documentation}
    \fi
    \label{sf:supplementalMaterial}}

\section{Acknowledgement}

This research was supported by SolidLab Vlaanderen (Flemish Government, EWI RRF project VV023/10) and Serendipity Engine (Research Foundation - Flanders (FWO) grant number S006323N).
Ruben Taelman is a postdoctoral fellow of the Research Foundation – Flanders (FWO) (1202124N).

\printbibliography
\addcontentsline{toc}{section}{References}

\onecolumn

\section*{Appendix}\label{sec:appendix}
\addcontentsline{toc}{section}{Appendix}
\FloatBarrier

\begin{figure*}[h]
    \centering
    \includegraphics[width=0.70\linewidth]{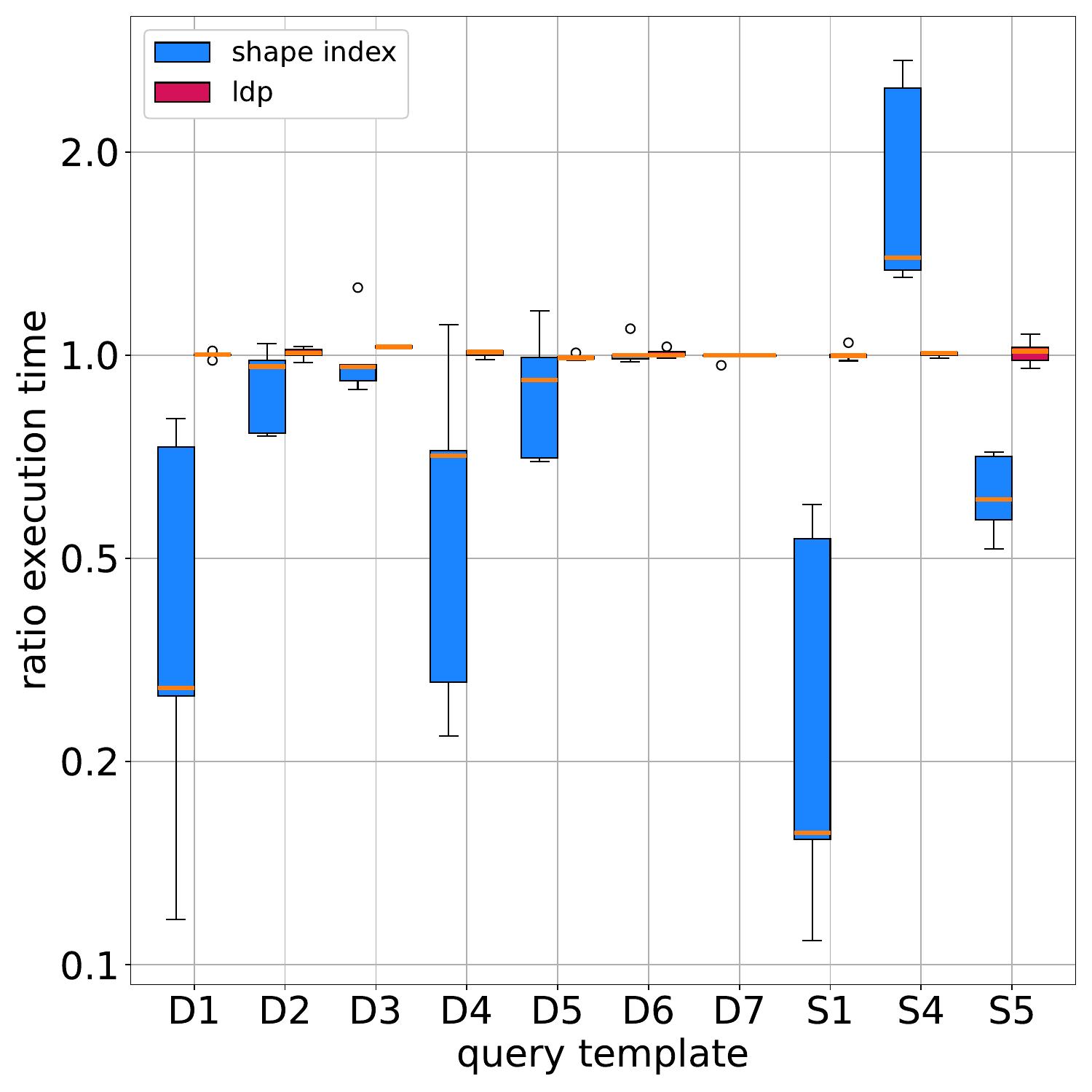}
    \caption{
    Comparison of query execution time with the type index approach.
    The ratio represents how the execution time of each method compares to that of the type index. 
    A ratio greater than $1$ indicates a slower execution time (\textbf{lower is better}).
    The shape index approach performs similarly or better than the other methods, except in the case of S4.
    }
    \label{fig:ratioApproaches}
\end{figure*}

\begin{table}
	\begin{center}
		\begin{tabular}{|l|l|l|l|}
			\hline
			query template & relation execution time & p-value & average ratio HTTP request \\
			\hline
			D1 & lesser & 1.14E-36 & 0.57 \\
			\hline
			D2 & lesser & 4.42E-04 & 0.88 \\
			\hline
			D3 & similar & 7.47E-01 (RH) & 0.97 \\
			\hline
			D4 & lesser & 2.07E-17 & 0.65 \\
			\hline
			D5 & lesser & 5.58E-03 & 0.88 \\
			\hline
			D6 & similar & 2.56E-01 (RH) & 1.12 \\
			\hline
			D7 & similar & 7.83E-01 (RH) & 1.12 \\
			\hline
			S1 & lesser & 1.12E-83 & 0.33 \\
			\hline
			S4 & greater & 3.76E-22 & 13.00 \\
			\hline
			S5 & lesser & 3.12E-17 & 0.44 \\
			\hline
		\end{tabular}
	\end{center}
	\caption{Table comparing the shape index approach to the state-of-the-art. RH, indicate that the p-value is associated to the rejected hypothesis. Every query performs better or similarly to the state-of-the-art with the shape index approach except for interactive-short-4. One might expect the average ratio of HTTP requests for D6 and D7 to be one. 
In our implementation, however, the query subsumption algorithm naively retrieved nested shapes, 
leading to some shapes being dereferenced multiple times. 
This is an implementation artifact rather than an inherent property of the method. 
Even in this worst case, the impact on the number of HTTP requests is small and does not affect the overall conclusions, 
especially given that requests are sent in parallel.}
	\label{tab:statSignificanceStateOfTheArt}
\end{table}

\renewcommand{\arraystretch}{1.1}
\begin{table}[h]
    \centering
    \resizebox{\textwidth}{!}{
    \begin{tabular}{|c|c|c|c||c|c|c||c|c|c|}
        \hline
        \textbf{Template} 
        & \multicolumn{3}{|c||}{\textbf{Shape Index}} 
        & \multicolumn{3}{c||}{\textbf{Type Index}} 
        & \multicolumn{3}{c|}{\textbf{LDP}} \\
        \hline
        & \textbf{IT ($n \times ms$)} & \textbf{SFT ($n \times ms$)} & \textbf{LRT ($n \times ms$)} 
        & \textbf{IT ($n \times ms$)} & \textbf{SFT ($n \times ms$)} & \textbf{LRT ($n \times ms$)} 
        & \textbf{IT ($n \times ms$)} & \textbf{SFT ($n \times ms$)} & \textbf{LRT ($n \times ms$)} \\
        \hline
        D1 & $0.00$ & $\boldsymbol{1.70}$ & $\boldsymbol{16.96}$ & $0.00$ & $0.22$ & $11.22$ & $0.00$ & $0.21$ & $10.67$ \\
        \hline
        D2 & $0.00$ & $\boldsymbol{1.91}$ & $119.54$ & $0.00$ & $0.31$ & $119.75$ & $0.00$ & $0.23$ & $\boldsymbol{124.94}$ \\
        \hline
        D3 & $0.00$ & $0.00$ & $85.14$ & $0.00$ & $0.00$ & $\boldsymbol{97.01}$ & $0.00$ & $0.00$ & $85.78$ \\
        \hline
        D4 & $0.00$ & $\boldsymbol{0.36}$ & $\boldsymbol{10.25}$ & $0.00$ & $0.06$ & $1.92$ & $0.00$ & $0.07$ & $1.75$ \\
        \hline
        D5 & $0.00$ & $\boldsymbol{1.03}$ & $\boldsymbol{13.82}$ & $0.00$ & $0.74$ & $12.18$ & $0.00$ & $0.74$ & $12.16$ \\
        \hline
        D6 & $0.00$ & $\boldsymbol{0.08}$ & $16.08$ & $0.00$ & $0.08$ & $\boldsymbol{20.04}$ & $0.00$ & $0.08$ & $18.88$ \\
        \hline
        D7 & $0.00$ & $\boldsymbol{0.08}$ & $\boldsymbol{0.28}$ & $0.00$ & $0.07$ & $0.20$ & $0.00$ & $0.07$ & $0.21$ \\
        \hline
        S1 & $0.00$ & $\boldsymbol{0.68}$ & $\boldsymbol{1.25}$ & $0.00$ & $0.18$ & $0.28$ & $0.00$ & $0.18$ & $0.27$ \\
        \hline
        S4 & $0.00$ & $0.49$ & $0.00$ & $\boldsymbol{0.01}$ & $\boldsymbol{0.66}$ & $\boldsymbol{0.04}$ & $0.01$ & $0.66$ & $0.04$ \\
        \hline
        S5 & $0.00$ & $0.25$ & $0.05$ & $0.00$ & $0.34$ & $0.10$ & $0.00$ & $\boldsymbol{0.37}$ & $\boldsymbol{0.13}$ \\
        \hline
        S7 & $\boldsymbol{0.00}$ & $\boldsymbol{0.00}$ & $\boldsymbol{0.00}$ & - & - & - & - & - & - \\
        \hline
    \end{tabular}
    }
    \caption{Table showing the average diefficiency metric for different approaches across query templates, measured at three time thresholds: instantaneous time (IT, 0.1s), seamless flow time (SFT, 1s), and at the time of the last result (LRT).
    The diefficiency is expressed as the number of results (n) per millisecond (ms).
    For most queries, diefficiency at IT is 0, as no results are produced quickly enough.
    At SFT, the shape index approach generally yields better performance, while at LRT, results tend to converge across approaches.}
	\label{tab:dief}
\end{table}

\renewcommand{\arraystretch}{1.1}
\begin{table}[h]
    \centering
    \resizebox{\textwidth}{!}{
    \begin{tabular}{|c|c|c|c||c|c|c||c|c|c|}
        \hline
        \textbf{Template} 
        & \multicolumn{3}{|c||}{\textbf{Shape Index}} 
        & \multicolumn{3}{c||}{\textbf{Type Index}} 
        & \multicolumn{3}{c|}{\textbf{LDP}} \\
        \hline
        & \textbf{FT ($s$)} & \textbf{TT ($s$)} & \textbf{WT ($s$)} 
        & \textbf{FT ($s$)} & \textbf{TT ($s$)} & \textbf{WT ($s$)} 
        & \textbf{FT ($s$)} & \textbf{TT ($s$)} & \textbf{WT ($s$)} \\
        \hline
        D1 & $\boldsymbol{0.66}$ & $\boldsymbol{0.11}$ & $0.23$ & $1.51$ & $1.09$ & $\boldsymbol{0.21}$ & $1.55$ & $1.02$ & $0.21$ \\
        \hline
        D2 & $\boldsymbol{1.25}$ & $0.00$ & $0.78$ & $1.91$ & $\boldsymbol{0.00}$ & $0.39$ & $1.88$ & $0.00$ & $\boldsymbol{0.38}$ \\
        \hline
        D3 & $3.74$ & $\boldsymbol{0.00}$ & $0.00$ & $\boldsymbol{3.67}$ & $0.00$ & $0.00$ & $3.77$ & $0.00$ & $0.00$ \\
        \hline
        D4 & $\boldsymbol{1.74}$ & $0.00$ & $0.00$ & $3.14$ & $0.00$ & $0.00$ & $3.17$ & $\boldsymbol{0.00}$ & $0.00$ \\
        \hline
        D5 & $\boldsymbol{0.49}$ & $\boldsymbol{0.03}$ & $0.22$ & $1.17$ & $0.04$ & $0.01$ & $1.18$ & $0.04$ & $\boldsymbol{0.00}$ \\
        \hline
        D6 & $1.42$ & $\boldsymbol{0.02}$ & $0.58$ & $\boldsymbol{1.40}$ & $0.04$ & $\boldsymbol{0.49}$ & $1.45$ & $0.07$ & $0.49$ \\
        \hline
        D7 & $\boldsymbol{1.48}$ & $1.19$ & $0.00$ & $1.56$ & $1.12$ & $0.00$ & $1.55$ & $\boldsymbol{1.11}$ & $0.00$ \\
        \hline
        S1 & $\boldsymbol{0.32}$ & $\boldsymbol{0.00}$ & $0.00$ & $1.30$ & $0.09$ & $0.00$ & $1.31$ & $0.09$ & $0.00$ \\
        \hline
        S4 & $0.51$ & $\boldsymbol{0.01}$ & $0.00$ & $\boldsymbol{0.34}$ & $0.01$ & $0.00$ & $0.34$ & $0.01$ & $0.00$ \\
        \hline
        S5 & $0.91$ & $\boldsymbol{0.00}$ & $0.00$ & $0.78$ & $0.68$ & $0.00$ & $\boldsymbol{0.75}$ & $0.76$ & $0.00$ \\
        \hline
        S7 & - & - & $\boldsymbol{0.00}$ & - & - & - & - & - & - \\
        \hline
    \end{tabular}
    }
    \caption{Continuous performance metrics showing First Result Arrival Time (FT), Termination Time (TT), and Waiting Time (WT). FT refers to the time taken for the first result to be delivered.
     TT captures the delay between the last result and the termination of the query engine.
    WT represents the total time the user waited more than one second between consecutive results, indicating interruptions in the seamless flow experience.}
	\label{tab:continuousPerf}
\end{table}

\begin{table}
	\begin{center}
		\begin{tabular}{|l|l|l|l|}
			\hline
			Query Template & avg (ms) & std (ms) & max (ms) \\
			\hline
			D1 & 0.062 & 0.073 & 0.539 \\
			\hline
			D2 & 0.106 & 0.461 & 4.655 \\
			\hline
			D3 & 0.040 & 0.034 & 0.133 \\
			\hline
			D4 & 0.043 & 0.068 & 0.672 \\
			\hline
			D5 & 0.015 & 0.015 & 0.025 \\
			\hline
			D6 & 0.068 & 0.039 & 0.174 \\
			\hline
			D7 & 0.041 & 0.101 & 1.005 \\
			\hline
			D8 & 0.033 & 0.045 & 0.426 \\
			\hline
			S1 & 0.074 & 0.064 & 0.560 \\
			\hline
			S2 & 0.309 & 0.327 & 2.791 \\
			\hline
			S3 & 0.072 & 0.069 & 0.650 \\
			\hline
			S4 & 0.042 & 0.032 & 0.144 \\
			\hline
			S5 & 0.043 & 0.057 & 0.551 \\
			\hline
			S6 & 0.132 & 0.343 & 3.473 \\
			\hline
			S7 & 0.176 & 0.158 & 0.971 \\
			\hline
		\end{tabular}
	\end{center}
	\caption{Query-Shape containment computation time (100 samples) is negligeable with the most restrictive shapes of our experiments.}
	\label{tab:queryShapeContainmentEval}
\end{table}

\begin{table}[h]
    \centering
    \ifreview\color{revtwo}\fi
    \begin{tabular}{|c|l|c|}
        \hline
        \textbf{Template} & \textbf{Description} & \textbf{Data-model selective} \\
        \hline
        D1 & All posts created by a person, with their content and creation date. & Yes \\
        \hline
        D2 & All messages (posts and comments) created by a person, with their content and creation date. & No \\
        \hline
        D3 & The tags attached to a person's messages, ranked by how often each occurs. & No \\
        \hline
        D4 & The locations of a person's comments, ranked by how often each occurs. & Yes \\
        \hline
        D5 & The distinct IP addresses a person has sent messages from. & No \\
        \hline
        D6 & The forums that contain a message created by a person. & No \\
        \hline
        D7 & The moderators of the forums that contain a message created by a person. & No \\
        \hline
        S1 & The profile of a person (name, gender, birthday, location, browser, creation date). & Yes \\
        \hline
        S3 & The friends of a person and the date each friendship was created. & Yes \\
        \hline
        S4 & The content (or image) and creation date of a message. & No \\
        \hline
        S5 & The creator of a message. & No \\
        \hline
        S7 & The comments that directly reply to a message, with their authors. & Yes \\
        \hline
    \end{tabular}
    \caption{\rtwo{Description of the query templates used in the evaluation. A \emph{message} denotes either a post or a comment; the words \emph{post} and \emph{comment} are used when a template targets that type specifically.}}
    \label{tab:queryTemplateDescription}
\end{table}
\FloatBarrier

\begin{Rtwo}
\begin{lstlisting}[style=sparql,caption={The S4 (interactive-short-4) query template.},label={lst:s4}]
PREFIX rdf: <http://www.w3.org/1999/02/22-rdf-syntax-ns#>
PREFIX rdfs: <http://www.w3.org/2000/01/rdf-schema#>
PREFIX xsd: <http://www.w3.org/2001/XMLSchema#>
PREFIX sn: <http://localhost:3000/www.ldbc.eu/ldbc_socialnet/1.0/data/>
PREFIX snvoc: <http://localhost:3000/www.ldbc.eu/ldbc_socialnet/1.0/vocabulary/>
PREFIX sntag: <http://localhost:3000/www.ldbc.eu/ldbc_socialnet/1.0/tag/>
PREFIX foaf: <http://xmlns.com/foaf/0.1/>
PREFIX dbpedia: <http://localhost:3000/dbpedia.org/resource/>
PREFIX dbpedia-owl: <http://localhost:3000/dbpedia.org/ontology/>

SELECT
    ?messageCreationDate
    ?messageContent
WHERE
{
    ?message snvoc:id ?messageId .
    ?message snvoc:creationDate ?messageCreationDate .
    ?message snvoc:content|snvoc:imageFile ?messageContent .
}
\end{lstlisting}
\end{Rtwo}

\begin{Rtwo}
\begin{lstlisting}[style=sparql,caption={The D6 (interactive-discover-6) query template.},label={lst:d6}]
PREFIX rdf: <http://www.w3.org/1999/02/22-rdf-syntax-ns#>
PREFIX rdfs: <http://www.w3.org/2000/01/rdf-schema#>
PREFIX xsd: <http://www.w3.org/2001/XMLSchema#>
PREFIX sn: <http://localhost:3000/www.ldbc.eu/ldbc_socialnet/1.0/data/>
PREFIX snvoc: <http://localhost:3000/www.ldbc.eu/ldbc_socialnet/1.0/vocabulary/>
PREFIX sntag: <http://localhost:3000/www.ldbc.eu/ldbc_socialnet/1.0/tag/>
PREFIX foaf: <http://xmlns.com/foaf/0.1/>
PREFIX dbpedia: <http://localhost:3000/dbpedia.org/resource/>
PREFIX dbpedia-owl: <http://localhost:3000/dbpedia.org/ontology/>

SELECT
DISTINCT
    ?forumId
    ?forumTitle
WHERE
{
    ?message snvoc:hasCreator ?person.
    ?forum snvoc:containerOf ?message;
            snvoc:id ?forumId;
            snvoc:title ?forumTitle.
}
\end{lstlisting}
\end{Rtwo}

\begin{Rtwo}
\begin{lstlisting}[style=sparql,caption={The D7 (interactive-discover-7) query template.},label={lst:d7}]
PREFIX rdf: <http://www.w3.org/1999/02/22-rdf-syntax-ns#>
PREFIX rdfs: <http://www.w3.org/2000/01/rdf-schema#>
PREFIX xsd: <http://www.w3.org/2001/XMLSchema#>
PREFIX sn: <http://localhost:3000/www.ldbc.eu/ldbc_socialnet/1.0/data/>
PREFIX snvoc: <http://localhost:3000/www.ldbc.eu/ldbc_socialnet/1.0/vocabulary/>
PREFIX sntag: <http://localhost:3000/www.ldbc.eu/ldbc_socialnet/1.0/tag/>
PREFIX foaf: <http://xmlns.com/foaf/0.1/>
PREFIX dbpedia: <http://localhost:3000/dbpedia.org/resource/>
PREFIX dbpedia-owl: <http://localhost:3000/dbpedia.org/ontology/>

SELECT
DISTINCT
    ?firstName
    ?lastName
WHERE
{
    ?message snvoc:hasCreator ?person.
    ?forum snvoc:containerOf ?message;
        snvoc:hasModerator ?moderator.
    ?moderator snvoc:firstName ?firstName .
    ?moderator snvoc:lastName ?lastName .
}
\end{lstlisting}
\end{Rtwo}
\end{document}